\documentstyle[12pt,epsf,axodraw]{article}
\newcommand{\beq}[1]{
\begin{equation}\label{#1}}
\newcommand{\eeq}{\end{equation}}
\newcommand{\bea}[1]{
\begin{eqnarray}\label{#1}}
\newcommand{\eea}{\end{eqnarray}}
\newcommand{\Prop}[1]
{
  \put(0,4){\line(1,0){#1}}
  \put(0,0){\line(0,1){4}}
  \put(#1,0){\line(0,1){4}}
}

\newcommand{\VKE}[3]
{
  \begin{picture}(#1,7)
  \put(#2,0){\Prop{#3}}
  \end{picture}
}

\begin{document}

\begin{titlepage}

\phantom{.}
\vspace{2cm}

\begin{center}
{\LARGE \bf The impact factor for the virtual photon \\
to light vector meson transition}

\vspace{1cm}

{\sc D.Yu.~Ivanov}${}^{1}$,
{\sc M.I.~Kotsky}${}^{2}$ and
{\sc A.~Papa}${}^{3}$
\\[0.5cm]
\vspace*{0.1cm}
${}^1$ {\it
Sobolev Institute of Mathematics, 630090 Novosibirsk, Russia 
                       } \\[0.2cm] 
\vspace*{0.1cm} 
${}^2${\it
Budker Institute for Nuclear Physics, 630090 Novosibirsk, Russia  \\
and Novosibirsk State University, 630090 Novosibirsk, Russia 
                       } \\[0.2cm]

\vspace*{0.1cm} ${}^3$ {\it Dipartimento di Fisica, Universit\`a
della Calabria \\
and Istituto Nazionale di Fisica Nucleare, Gruppo collegato di Cosenza \\
I-87036 Arcavacata di Rende, Cosenza, Italy
                       } \\[1.0cm]

\vspace*{0.5cm}

\centerline{\large \em \today}

\vskip2cm
{\bf Abstract\\[10pt]} \parbox[t]{\textwidth}{
We evaluate in the next-to-leading approximation the forward impact factor 
for the virtual photon to light vector meson transition in the case of 
longitudinal polarization. We find that in the hard kinematic domain, 
both in the leading and in the next-to-leading approximation, the expression 
for the impact factor factorizes, up to power suppressed corrections,
into the convolution of a perturbatively calculable hard-scattering amplitude and 
a meson twist-2 distribution amplitude.
}

\vskip1cm
\end{center}

\vspace*{1cm}

\end{titlepage}

\section{Introduction}

Diffractive processes in high-energy particle physics are of central
interest in present and forthcoming experiments. Among them particularly important
are those processes with a large typical momentum transfer $Q^2$, so called 
semi-hard processes, since they allow a theoretical description by perturbative QCD.
The most suitable approach for such description in the limit of high energy 
$\sqrt s$ is the BFKL approach~\cite{BFKL}, which has become widely known 
since the discovery at HERA of the sharp rise of the proton structure function 
for decreasing value of the Bjorken variable $x$ (see, for example,~\cite{CDR98}). 

In the BFKL approach the high energy scattering amplitudes are expressed  
in terms of the Green's function of two interacting Reggeized gluons and of 
the impact factors of the colliding particles~(see, for instance,~\cite{FF98}).
The BFKL equation allows to determine this Green's function. 
The impact factors depend on the process in question and must be calculated 
separately. 

The BFKL equation was initially derived in the leading logarithmic
approximation (LLA) which means summation of all terms of the type
$(\alpha_s\ln(s))^n$. The most important disadvantage of LLA is that neither
the scale of energy nor the argument of the QCD running coupling constant
$\alpha_s$ are constrained in this approximation. The corresponding
uncertainties related with the change of these scales diminish the
predictive power of the LLA and restrict its application to the
phenomenology. This is why the generalization of the BFKL approach to the 
next-to-leading logarithmic approximation (NLA), which means resummation  
of both the leading terms $(\alpha_s\ln(s))^n$ and of the terms 
$\alpha_s(\alpha_s\ln(s))^n$, is very important.  
The radiative corrections to the kernel of the BFKL equation were calculated 
several years ago~\cite{LF89,FRK95,FRK96,FL93,FFFKLQ,CCH} and the explicit 
form of the kernel of the equation in the NLA is known now~\cite{FL98,CC98} 
for the case of forward scattering. Instead, the problem of calculating NLA 
impact factors has been solved so far only for colliding 
partons~\cite{FFKP99,Cia}
and for forward jet production~\cite{Bartels:2002yj}, while no 
impact factors for colorless objects are known with the NLA accuracy.

It is clear that for a complete NLA description in the BFKL 
approach one needs to know the impact factors describing the transitions
between colorless particles, analogously as in the 
DGLAP approach one should know not only the parton distributions, but also 
the coefficient functions.

The impact factor for the $\gamma^* \to \gamma^*$ transition is certainly 
the most important one from the phenomenological point of view, since it would
open the way to predictions of the $\gamma^* \gamma^*$ total cross section. 
However, its calculation in the NLA turns out to be rather complicated and so far 
only partial results are available \cite{Bartels:2002uz}. For an
alternative approach to the calculation of the photon impact factor, see also
Ref.~\cite{Fadin:2002tu}.

In this paper we consider instead the NLA impact factor for the transition from a
virtual photon $\gamma^*$ to a light neutral vector meson $V=\rho^0, \omega, \phi$. 
As we will explain in detail below, the evaluation of this impact factor is 
considerably simpler than for the case of the $\gamma^*\to \gamma^*$ impact factor.
Indeed, we obtained a closed analytical expression for this impact factor 
in the NLA. This result should help to understand more easily the main physical 
effects of the radiative corrections in the BFKL approach. As a matter of fact, 
the knowledge of the $\gamma^* \to V$ impact factor allows for the first time 
to determine completely within perturbative QCD and with NLA accuracy the amplitude 
of a physical process, the $\gamma^* \gamma^* \to V V$ reaction. This possibility 
is interesting not only for theoretical reasons, since it could shed light on the 
correct choice of energy scales in the BFKL approach and could be used to compare 
different approaches such as BFKL and DGLAP, but also for the possible applications 
to the phenomenology. Indeed, the calculation of the 
$\gamma^* \to V$ impact factor is the first step towards the application of BFKL 
approach to the description of processes such as the vector meson electroproduction
$\gamma^* p\to V p$, being carried out at the HERA collider, and the production 
of two mesons in the photon collision $\gamma^*\gamma^*\to VV$ 
or $\gamma^* \gamma \to VJ/\Psi$, which can be studied at high-energy $e^+e^-$ 
and $e\gamma$ colliders.

The paper is organized as follows. In the next Section we will recall
the definition of impact factor in the BFKL approach and outline the general 
framework of the calculation; in Section~3 we give a detailed derivation of 
the leading logarithmic approximation (LLA) impact factor; Section~4 is 
devoted to the calculation of the impact factor in the NLA; in Section~5 we 
summarize our results and draw some conclusions.

\section{General framework}

In the BFKL approach (see, for example, Ref.~\cite{FF98} for details) 
the scattering amplitude $\left({\cal A} \right)_{AB}^{A^\prime
B^\prime}$ of the process $AB \rightarrow A^\prime B^\prime$, were $A, B, 
A^\prime, B^\prime$ are colorless particles, for c.m.s. energy $\sqrt{s}  
\rightarrow \infty$ and fixed momentum transfer $\Delta \approx \Delta_\perp$ 
($\perp$ means transverse to the initial particle momenta plane) is expressed in 
terms of the Mellin transform of the Green's function of two interacting Reggeized
gluons in the color singlet state $G_\omega$ and of the impact factors of the 
colliding particles $\Phi_{A\to A^\prime}$ and $\Phi_{B\to B^\prime}$:
$$
{\cal I}m_s\left( {\cal A} \right)_{AB}^{A^\prime B^\prime}
= \frac{s}{(2\pi)^{D-2}}\int\frac{d^{D-2}q_1}{\vec q_1^{\:2}(\vec q_1
- \vec \Delta)^2}\int\frac{d^{D-2}q_2}{\vec q_2^{\:2}(\vec q_2
- \vec \Delta)^2}
$$
\beq{11}
\times\Phi_{A\to A^\prime}(\vec q_1, \vec \Delta, s_0)
\int_{\delta - i\infty}^{\delta + i\infty}\frac{d\omega}{2\pi i}\left[
\left( \frac{s}{s_0} \right)^\omega G_\omega(\vec q_1, \vec q_2,
\vec \Delta) \right]\Phi_{B\to B^\prime}(-\vec q_2, -\vec \Delta, s_0),
\eeq
where ${\cal I}m_s$ means $s$-channel imaginary part, the vector sign 
is used for denotation of the transverse components, 
$D = 4 + 2\varepsilon$ is the space-time dimension different from $4$ 
to regularize both infrared and ultraviolet divergences and  $s_0$ is the
energy scale parameter. The Green's function obeys the BFKL equation 
generalized for non-zero momentum transfer~\cite{FF98}
\beq{12}
\omega \, G_\omega(\vec q_1, \vec q_2, \vec \Delta) = \vec q_1^{\:2}
(\vec q_1 - \vec \Delta)^2\delta^{(D-2)}(\vec q_1 - \vec q_2) + \int\frac
{d^{D-2}k}{\vec k^2(\vec k - \vec \Delta)^2}{\cal K}(\vec q_1,
\vec k, \vec \Delta)G_\omega(\vec k, \vec q_2, \vec \Delta)\;,
\eeq
where ${\cal K}$ is the NLA singlet kernel, and is completely defined 
by this equation. The definition of the NLA impact factors has been given 
in Ref.~\cite{FF98}; in the case of scattering of the particle $A$ off 
a Reggeized gluon with momentum $q_1$, for transverse momentum 
$\vec \Delta$ and singlet color representation in the $t$-channel, the 
impact factor has the form~\cite{FFKP99}
$$ \Phi_{A\to A^\prime}(\vec q_1, \vec \Delta, s_0)
= \frac{\delta^{c c'}}{\sqrt{N_c^2-1}} 
\left[\left( \frac{s_0}{\vec q_1^{\:2}} \right)^{\frac{1}{2}
\omega(-\vec q_1^{\:2})}\left( \frac{s_0}{(\vec q_1 - \vec \Delta)^2}
\right)^{\frac{1}{2}\omega(-(\vec q_1 - \vec \Delta) ^2)}\right.
$$
$$
\left.
\times\sum_{\{f\}}\int\frac{d\kappa}{2\pi}\theta(s_\Lambda - \kappa)
d\rho_f\Gamma^c_{A\{f\}}\left( \Gamma^{c^\prime}_{A^\prime\{f\}} \right)^*\right]
$$
\beq{13}
- \frac{1}{2}\int\frac{d^{D-2}k}{\vec k^2(\vec k - \vec \Delta)^2}
\Phi_{A\to A^\prime}^{Born}(\vec k, \vec \Delta, s_0){\cal K}_r^{Born}
(\vec k,\vec q_1, \vec \Delta)\ln\left( \frac{s_\Lambda^2}{s_0(\vec k - \vec
q_1)^2} \right)\;,
\eeq
where $c$, $c^\prime$ are the color indices of the Reggeized gluon in the initial
and in the final state, respectively, $\omega(t)$ is the Reggeized gluon trajectory 
in the LLA and the intermediate parameter $s_\Lambda$ must be taken tending to 
infinity. The integration in the first term of the above equality is carried out 
over the phase space $d\rho_f$ and over the squared invariant mass $\kappa$ of 
the system $\{f\}$ produced in the fragmentation region of the particle $A$, 
$\Gamma^c_{A\{f\}}$ are the particle-Reggeon effective vertices for 
this production and the sum is taken over all systems $\{f\}$ which 
can be produced in the NLA. The second term in Eq.~(\ref{13}) is the 
counterterm for the LLA part of the first one, so that the logarithmic 
dependence of both terms on the intermediate parameter $s_\Lambda \rightarrow 
\infty$ disappears in their sum; ${\cal K}_r^{Born}$ is the part of the leading 
order BFKL kernel related to the real gluon production (see Ref.~\cite{FF98} 
for more details). It was shown in Ref.~\cite{FM99} that the definition~(\ref{13}) 
guarantees the infrared finiteness of the impact factors of colorless particles. 


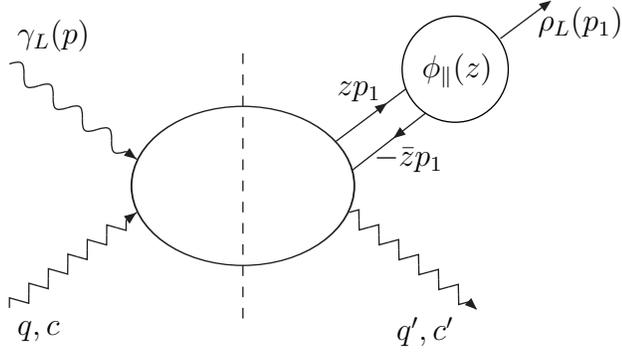
\begin{figure}[tb]
\begin{center}
\begin{picture}(200,120)(-20,-20)

\Photon(-8,86)(36,53){3}{4}
\ArrowLine(36,53)(40,50)
\ArrowLine(115,56.75)(150,83)
\ArrowLine(156,73)(121,45.75)
\GCirc(160,84){20}{1}
\ZigZag(-8,-6)(36,27){3}{6}
\ArrowLine(36,27)(40,30)
\Oval(80,40)(30,42)(0)
\ZigZag(120,30)(165,-3.75){3}{6}
\ArrowLine(165,-3.75)(168,-6.5)
\DashLine(80,90)(80,-10){4}
\LongArrow(177,95.55)(195,109)
\Text(-5,97)[l]{$\gamma_{L}(p)$}
\Text(-5,-15)[l]{$q,c$}
\Text(160,-15)[r]{$q^{\prime},c^{\prime}$}
\Text(116,76)[l]{$zp_1$}
\Text(130,50)[l]{$-{\bar z}p_1$}
\Text(148,84)[l]{$\phi_{\parallel}(z)$}
\Text(192,102)[l]{$\rho_L(p_1)$}

\end{picture}
\end{center}
\caption[]{The kinematics of the virtual photon to vector meson 
impact factor.}
\end{figure}


In this paper we will study in the NLA and in the forward ($\vec \Delta=0$) case
the impact factor which describes the transition of a virtual photon to a 
light neutral meson $\Phi_{\gamma^*\to V}$, $V=\rho^0, \omega, \phi$.
To start with, let us describe the kinematics which is presented in Fig.~1. We 
introduce two auxiliary Sudakov vectors $p_1$ and $p_2$, such that $p_1^2=p_2^2=0$ 
and $2 (p_1p_2)=s$. The virtual photon momentum is $p = p_1-\zeta p_2$,
where $\zeta=Q^2/s$. The large negative virtuality of the photon $p^2=-Q^2$ 
introduces a hard scale in the problem and justifies the application of 
perturbative QCD. We denote the Reggeon momenta as
\bea{reggeonM}
&&
q=\frac{\kappa +Q^2+\vec q^{\:2}}{s}p_2+q_\perp \, , \quad q^2=q^2_\perp=
-\vec q^{\:2} \, , \nonumber \\
&&
q^\prime=\frac{\kappa +\vec q^{\:2}}{s}p_2+q_\perp \, , 
\eea
where $\kappa$ is the squared invariant mass of the intermediate state
produced in the virtual photon-Reggeon interaction. The component of the    
Reggeon momentum proportional to the Sudakov vector $p_1$ is inessential for 
the analysis of the impact factor, therefore it is not included in 
Eqs.~(\ref{reggeonM}). We restrict our consideration to the forward case, 
when the momentum transfer vector has only the longitudinal component, 
$q-q^\prime =\zeta p_2$. We remark that our denotation of Reggeon momenta here
is slightly different from Eq.~(\ref{13}), since we use the symbols $q$ 
and $q^\prime$ for the momenta of the incoming and outgoing Reggeon, 
instead of $q_1$ and $q_2$ of Eq.~(\ref{13}). 
We assume that both the square of the Reggeon transverse momentum 
$\vec q^{\, \, 2}$ and the virtuality of the photon $Q^2$ are much larger than any 
hadronic scale. In what follows we will neglect all power suppressed
contributions. Therefore we neglect the vector meson mass and denote the    
momentum of the produced vector meson by the Sudakov vector $p_1$. 

We will show that in this kinematics the impact factor can be calculated in
the collinear factorization framework~\cite{earlyCZ,earlyBL,earlyER} which was 
developed for the QCD description of hard exclusive processes. 
We will demonstrate by explicit calculation that the dominant helicity 
amplitude is a transition of the longitudinally polarized photon $\gamma^*_L$ 
into the longitudinally polarized meson $V_L$, and that both in LLA and in 
NLA the expression for the impact factor factorizes into the 
convolution\footnote{Here and in the following we consider the {\em unprojected}
impact factor, i.e. the impact factor as defined in Eq.~(\ref{13}) except
for the singlet color projector $\delta^{cc^\prime}/\sqrt{N_c^2-1}$.
The application of the color projector on (\ref{fact}) will result in the
replacement $\delta^{cc\prime}\to \sqrt{N_c^2-1}$.}
\beq{fact}
\Phi_{\gamma^*_L\to V_L}(\alpha,s_0)
=-\frac{4\pi e_q f_V \delta^{cc^\prime}}{N_c Q}
\int\limits^1_0\, dz\, T_H(z,\alpha,s_0,\mu_F, \mu_R) \, 
\phi_\parallel(z,\mu_F)  
\eeq
of a perturbatively calculable hard-scattering amplitude, $T_H$, and 
a meson twist-2 distribution amplitude, $\phi_\parallel(z)$. The distribution
amplitude is defined as the vacuum-to-meson matrix element of the light-cone 
operator 
\beq{Vdampl}
\langle0|\bar \Psi(0) \gamma^\mu \Psi(y)|V_L(p_1)\rangle_{y^2 \to 0}        
 = f_V\, p_1^\mu
\int\limits^1_0 \!dz\, e^{-iz(p_1y)}\, \phi_\parallel (z,\mu_F)\, ,
\eeq
where $f_V\simeq 200\, \rm{MeV}$ is the meson decay constant; it has the meaning 
of the probability amplitude to find a meson in a state with minimal number 
of constituents -- quark and antiquark -- when they are separated by
small transverse distances, $r_\bot\sim 1/\mu_F$. Here $\mu_F^2\sim
Q^2,\vec q^{\:2}$ is a factorization (or separation) scale at which soft and
hard physics factorizes according to Eq.~(\ref{fact}). The variable $z$ 
corresponds to the longitudinal momentum fraction carried by the quark, for 
the antiquark the fraction is $\bar z=1-z$. Finally, $N_c=3$ in 
Eq.~(\ref{fact}) stands for the number of QCD colors, we introduced the ratio 
\beq{ratio}
\alpha = \frac{\vec q^{\:2}}{Q^2} \ ,
\eeq
and separated other factors for further convenience. We use the
conventions of~\cite{Braun} for the vector meson distribution amplitude.

Eqs.~(\ref{fact}) and (\ref{Vdampl}) give the contribution to the impact factor
of the single light quark flavor $q$ carrying electric charge $e_q$. According 
to the neutral mesons flavor structure,
\beq{flstructure}
|\rho^0\rangle =\frac{1}{\sqrt{2}}
\left(|\bar u u\rangle -|\bar d d\rangle\right)\, ,
\quad 
|\omega\rangle =\frac{1}{\sqrt{2}}
\left(|\bar u u\rangle +|\bar d d\rangle\right)\, , 
\quad
|\phi\rangle =|\bar s s\rangle \, , 
\eeq
$e_q$ should be replaced by 
$e/\sqrt{2}$, $e/(3\sqrt{2})$ and $-e/3$ for the case of $\rho^0$, $\omega$ and
$\phi$ meson production, respectively.

The hard-scattering amplitude $T_H$ is represented as a series in the QCD 
running coupling constant $\alpha_S(\mu_R)$, where the renormalization scale 
$\mu_R$ is of the order of the hard scale $\mu_R^2\sim Q^2,\vec q^{\:2}$. 
The distribution amplitude is a nonperturbative function, but its dependence on
the scale $\mu_F$ is perturbative,
\beq{evEq}
\mu_F^2\frac{d\phi_\parallel (z,\mu_F)}{d\mu_F^2}=\int\limits^1_0 
V(z,z^\prime)\phi_\parallel (z^\prime,\mu_F) dz^\prime
\, ,
\eeq
and is determined by the renormalization group equation for the nonlocal    
light-cone operator in the left-hand side of Eq.~(\ref{Vdampl}). It is 
renormalized at the scale $\mu_F$, so that the distribution amplitude depends 
on $\mu_F$ as well. The insertion of the path-ordered gauge factor between the
quark field operators in Eq.~(\ref{Vdampl}), which restores the gauge 
invariance of the nonlocal matrix element, is implied. The local limit of 
this operator corresponds to the conserved vector current, therefore  
\begin{equation}
\int\limits^1_0 dz\, \phi_\parallel (z,\mu_F)\, 
\eeq
is a renormalization-scale-invariant 
quantity and the normalization condition for the
distribution amplitude
\beq{norm}
\int\limits^1_0 dz\, \phi_\parallel (z,\mu_F)=1 \, 
\eeq
corresponds to our convention for $f_V$. The kernel of the evolution
equation (\ref{evEq}) is known as an expansion in $\alpha_S(\mu_F)$ 
up to the second order
\beq{kern}
V(z,z^\prime)=\frac{\alpha_s(\mu_F)}{2\pi}V^{(1)}(z,z^\prime)+
\left(\frac{\alpha_s(\mu_F)}{2\pi}\right)^2V^{(2)}(z,z^\prime)+\dots \, .
\eeq  
For the purposes of the present work, we need to remind the expression
for $V^{(1)}$,
\beq{V(1)}
V^{(1)}(z,z^\prime)=C_F\,
\left[\frac{1-z}{1-z^\prime}\left(1+\frac{1}{z-z^\prime}\right)\theta(z-z^\prime)
+\frac{z}{z^\prime}\left(1+\frac{1}{z^\prime-z}\right)\theta(z^\prime-z)\right]_+\;,
\eeq
where
\begin{equation}
[f(z,z^\prime)]_+ \equiv f(z,z^\prime)-\delta(z-z^\prime) \int_0^1 dt \,  
f(t,z)\;.
\end{equation}

Let us now discuss briefly the cancellation of divergences in
the final expression for the NLA impact factor. We perform the calculations with 
unrenormalized quantities, the bare strong 
coupling constant $\alpha_S$ and the bare meson distribution amplitude 
$\phi_\parallel^{(0)} (z)$. Therefore the NLA expression for the 
hard-scattering amplitude $T_H$ expressed in terms of these quantities
will contain both ultraviolet and infrared divergences, appearing as poles 
in the common dimensional regularization parameter $\varepsilon$. 
The ultraviolet divergences will disappear after the strong coupling constant 
renormalization. In the $\overline{\rm{MS}}$ scheme, it is given 
with the required accuracy by 
\beq{alsrem}
\alpha_s=\alpha_s(\mu_R)\left[1+
\frac{\alpha_s(\mu_R)}{4\pi}\beta_0\left(
\frac{1}{\hat \varepsilon}+\ln\left(\frac{\mu_R^2}{\mu^2}\right)\right)\right]\, ,
\eeq
where $\mu$ is the dimensional parameter introduced by the dimensional 
regularization, 
\beq{beta0}
\beta_0=\frac{11\,N_c}{3}-\frac{2n_f}{3}\, , \;\;\;\;\; \frac{1}{\hat
\varepsilon}=\frac{1}{\varepsilon}+\gamma_E-\ln(4\pi)\;,
\eeq
where $n_f$ is the effective number of light quark flavors, $\gamma_E$ is 
Euler's constant. The surviving infrared divergences will be only due
to collinear singularities, the soft singularities cancel as usual after  
summing the ``virtual'' and the ``real'' parts of the radiative corrections.
Since impact factors should be infrared-finite objects for physical transitions,
it must be possible to absorb the remaining infrared divergences into the 
definition of the nonperturbative distribution amplitude. We will show that 
this is achieved by the substitution of the bare distribution amplitude 
by the renormalized one, given in the $\overline{\rm{MS}}$ scheme by
\beq{renampl}
\phi_\parallel^{(0)} (z) \rightarrow 
\phi_\parallel (z,\mu_F) -\frac{\alpha_s(\mu_F)}{2\pi}
\left(\frac{1}{\hat \varepsilon}+\ln\left(\frac{\mu_F^2}{\mu^2}\right)\right)
\int\limits^1_0 V^{(1)}(z,z^\prime)\phi_\parallel (z^\prime,\mu_F) 
dz^\prime\; ,
\eeq
which leads to a finite expression for $T_H$ in the NLA.
The success of the procedure described above and the finiteness 
of the $z$ integral obtained in the factorization formula~(\ref{fact}) 
would mean that the NLA impact factor in question can be unambiguously
calculated in the collinear factorization approach.  


\begin{figure}[tb]
\begin{center}
\begin{picture}(400,200)(0,0)

\CArc(50,150)(15,90,270)
\ArrowLine(50,165)(110,165)
\ArrowLine(80,135)(50,135)
\ArrowLine(110,135)(80,135)
\Photon(0,150)(28,150){3}{4}
\ArrowLine(28,150)(35,150)
\ZigZag(50,100)(50,128){3}{6}
\ArrowLine(50,128)(50,135)
\ZigZag(80,135)(80,107){3}{6}
\ArrowLine(80,107)(80,100)
\GCirc(120,150){20}{1}
\Line(140,150)(155,150)
\ArrowLine(155,150)(160,150)

\Text(0,160)[l]{$p$}
\Text(40,105)[c]{$q$}
\Text(90,105)[c]{$q'$}
\Text(160,160)[c]{$p_1$}
\Text(120,150)[c]{$V$}

\Text(140,110)[c]{$(a)$}

\CArc(250,150)(15,90,270)
\ArrowLine(250,165)(280,165)
\ArrowLine(280,165)(310,165)
\ArrowLine(310,135)(250,135)
\Photon(200,150)(228,150){3}{4}
\ArrowLine(228,150)(235,150)
\ZigZag(250,100)(250,128){3}{6}
\ArrowLine(250,128)(250,135)
\ZigZag(280,165)(280,140){3}{6}
\ZigZag(280,130)(280,107){3}{6}
\ArrowLine(280,107)(280,100)
\GCirc(320,150){20}{1}
\Line(340,150)(355,150)
\ArrowLine(355,150)(360,150)

\Text(340,110)[c]{$(b)$}

\CArc(50,50)(15,90,270)
\ArrowLine(50,65)(110,65)
\ArrowLine(80,35)(50,35)
\ArrowLine(110,35)(80,35)
\Photon(0,50)(28,50){3}{4}
\ArrowLine(28,50)(35,50)
\ZigZag(50,0)(50,30){3}{6}
\ZigZag(50,40)(50,58){3}{5}
\ArrowLine(50,58)(50,65)
\ZigZag(80,35)(80,7){3}{6}
\ArrowLine(80,7)(80,0)
\GCirc(120,50){20}{1}
\Line(140,50)(155,50)
\ArrowLine(155,50)(160,50)

\Text(140,10)[c]{$(c)$}

\CArc(250,50)(15,90,270)
\ArrowLine(250,65)(280,65)
\ArrowLine(280,65)(310,65)
\ArrowLine(310,35)(250,35)
\Photon(200,50)(228,50){3}{4}
\ArrowLine(228,50)(235,50)
\ZigZag(250,0)(250,30){3}{6}
\ZigZag(250,40)(250,58){3}{5}
\ArrowLine(250,58)(250,65)
\ZigZag(280,65)(280,40){3}{6}
\ZigZag(280,30)(280,7){3}{6}
\ArrowLine(280,7)(280,0)
\GCirc(320,50){20}{1}
\Line(340,50)(355,50)
\ArrowLine(355,50)(360,50)

\Text(340,10)[c]{$(d)$}

\end{picture}
\end{center}
\caption[]{Feynman diagrams at the lowest order for the $\gamma^*R\to 
VR^\prime$ transition amplitude.}
\end{figure}
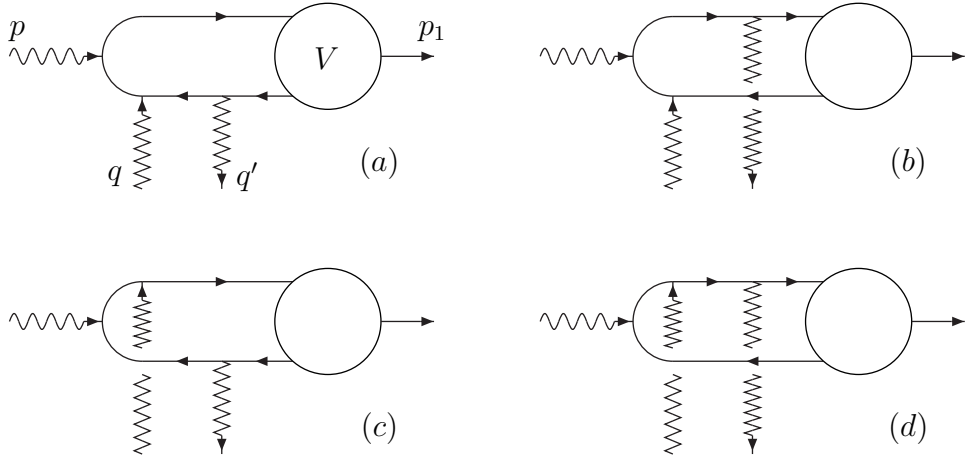

\section{Impact factor in LLA}

When the calculation of the impact factors in LLA is concerned,
Reggeons are equivalent to gluons with polarization vectors $p_2^\mu/s$
and the definition of the impact factor is reduced to the following integral of
the $\kappa$-channel discontinuity of the Reggeon (gluon) amplitude
\beq{Disc}
\Phi_{\gamma^*\to V}=-i\int \frac{d \kappa}{2\pi}\, {\it Disc}_{\kappa}\, 
{\cal A}_{\gamma^*R\to V\,R^\prime} \ ,
\eeq
where ${\cal A}_{\gamma^*R\to VR^\prime}$ is given by the Fourier transform 
of the matrix element of the electromagnetic current $J^\mu_{\rm{em}}=-e_q\bar 
\Psi\gamma^\mu \Psi$,
\beq{ampl}
(2\pi)^4\delta^4(p+q-p_1-q^\prime){\cal A}_{\gamma^*R\to VR^\prime}=
\int d^4 x \, e^{-i(px)} e_\mu
\langle V(p_1)R^\prime|J_{\rm{em}}^\mu (x)|R\rangle \, ,
\eeq
$e_\mu$ being the polarization vector of the virtual photon.

The lowest order Feynman diagrams for ${\cal A}_{\gamma^*R\to VR^\prime}$ 
having $\kappa$-channel discontinuity are shown in Fig.~2.
Let us consider the diagram $(a)$:
\bea{a}
&&
(2\pi)^4\delta^4(p+q-p_1-q^\prime){\cal A}^{(a)}_{\gamma^*R\to VR^\prime}=
-e_q(ig)^2\int d^4xd^4yd^4y_1e^{-i(px)-i(qy_1)+i(q^\prime y)}
 \nonumber \\
&&
\times \left(t^ct^{c^\prime}\right)_{ij}e_\mu
\langle V(p_1)|T\left[
\overline{  \Psi}^i (x)\gamma^\mu
\stackrel{\VKE{50}{2}{25}}{\Psi (x)
\overline{  \Psi} (y_1)}\frac{\not\! p_2}{s} \stackrel{\VKE{50}{2}{29}}
{ \Psi (y_1)
\overline{  \Psi} (y)}  \frac{\not\! p_2}{s} \Psi^j (y)
\right]|0\rangle \, ,
\eea
here the symbols $\stackrel{\VKE{50}{2}{25}}{\Psi (x)
\overline{  \Psi} (y_1)}$ and $\stackrel{\VKE{50}{2}{29}}
{ \Psi (y_1)\overline{  \Psi} (y)}$ stand for the fermion propagators, $g$ is 
QCD coupling, $\alpha_S=g^2/(4\pi)$, $t^c$ and $t^{c^\prime}$ denote the color
generators in the fundamental representation and $(i,j)$ are the quark
color indices. Using translational invariance, color neutrality of the meson
state and the Fierz identity, one can transform the meson matrix element to
the form
\bea{interm}
\langle V(p_1)|\overline{  \Psi}^i_\alpha (x) \Psi^j_\beta (y)|0\rangle &=&
e^{i(p_1y)}\langle V(p_1)|\overline{ \Psi}^i_\alpha (x-y) 
\Psi^j_\beta (0)|0\rangle \\
&=&
e^{i(p_1y)}\frac{\delta^{ij}}{N_c}\frac{1}{4}\left\{
(\gamma^\mu)_{\beta\alpha}
\langle V(p_1)|\overline{  \Psi} (x-y)\gamma_\mu \Psi (0)|0\rangle +\dots
\right\}\, , \nonumber
\eea
where ellipsis stand for the contributions of other Fierz structures. Fierz 
projectors having even number of $\gamma$ matrices correspond to the chiral-odd
configuration of the quark pair. Since in the massless limit 
chirality is conserved in the 
perturbative photon and gluon vertices, these chiral-odd structures do not 
contribute to the amplitude. Formally this appears as the vanishing of the 
trace of an odd number of $\gamma$ matrices in the quark loop. For the 
chiral-conserving case there is another possibility beyond $\gamma_\mu$,    
the Fierz structure $\gamma_\mu\gamma_5$. The light-cone expansion for 
this operator starts from twist-3, see~\cite{Braun}, therefore it does not 
contribute to the impact factor at leading twist. The $\gamma_\mu\gamma_5$ 
operator is important for the production of transversely polarized vector 
mesons. 
It contributes, together with the twist-3 term parameterizing the $\gamma_\mu$ 
operator, to the leading power asymptotics for the transversely polarized meson
impact factor. But, in comparison to the production of longitudinally
polarized meson, the transverse case contains an additional suppression factor 
$m_V/Q$, therefore we will concentrate in what follows on the leading power 
asymptotics of the impact factor, which is given by the production of a 
longitudinally polarized meson.
     
Substituting~(\ref{interm}) into Eq.~(\ref{a}) we obtain
\bea{int2}
{\cal A}^{(a)}_{\gamma^*R\to
V_LR^\prime}&=&-\frac{e_qg^2\delta^{cc^\prime}}{8N_c}\int
\frac{d^4l_1d^4l_2d^4(x-y)}{(2\pi)^4}\delta^4(l_1-l_2-q)e^{-i((p+l_1)(x-y))}
\nonumber \\
&\times&  {\rm Sp}\left[\not\! e \frac{\not l_1}{l_1^2+i\epsilon}
\frac{\not\! p_2}{s}\frac{\not l_2}{l_2^2+i\epsilon}\frac{\not\! p_2}{s}
\gamma^\mu\right]
\langle V_L(p_1)|\overline{\Psi} (x-y)\gamma_\mu \Psi (0)|0\rangle \, .
\eea
The leading power asymptotics of the integral (\ref{int2}) originates from the
region where the interquark separation goes to the light-cone, $(x-y)^2\to 0$.
Replacing the bilocal quark operator by its light-cone limit
\beq{int3}
\langle V_L(p_1)|\overline{ \Psi} (y)\gamma^\mu \Psi (0)|0\rangle_{y^2\to 0} =
f_V p_1^\mu \int\limits^1_0 dz\, e^{iz(p_1y)}\phi_\parallel(z) \ ,
\eeq
and performing the integrals, we find
\bea{int4}
{\cal A}^{(a)}_{\gamma^*R\to V_LR^\prime}
&=&-\frac{e_qg^2f_V\delta^{cc^\prime}}{8N_c}
\int\limits^1_0 dz\phi_\parallel(z) \int d^4l_1 \delta^4(p_1z-p-l_1)
\nonumber \\
&\times& {\rm Sp}\left[\not\! e \frac{\not
l_1}{l_1^2+i\epsilon}
\frac{\not\! p_2}{s}\frac{\not l_1-\not\! q}{(l_1-q)^2+i\epsilon}
\frac{\not\! p_2}{s}\not\! p_1\right] \, .
\eea
This consideration shows that the quark lines entering the meson blob in the
Fig.~2(a) belong actually to the light-cone operator~(\ref{int3}). The
delta function in~(\ref{int4}) means that the remaining part of the diagram 
which contributes to the hard-scattering amplitude is calculated for the
quark and antiquark momenta, $p_1z$ and $p_1\bar z$, being on-shell.
     
Going back from the Reggeon scattering amplitude to the impact factor, we see
that in LLA only one particle intermediate states contribute to the
discontinuity in Eq.~(\ref{Disc}). 
In the case of the diagram Fig.~2$(a)$ it is an antiquark cut   
\beq{int5}
\frac{i}{2\pi}{\it Disc}_\kappa \frac{1}{(l_1-q)^2+i\epsilon}=
\delta((l_1-q)^2)=\delta((p_1z-p-q)^2)=
\delta (\bar z \kappa - z\vec q^{\:2}) \ . 
\eeq
According to Eqs.~(\ref{int4}) and (\ref{int5}), the contribution 
of diagram Fig.~2$(a)$ to the impact factor reads
\beq{int6}
\Phi^{(a)}_{\gamma^*_L\to V_L}=-\frac{e_qg^2f_V\delta^{cc^\prime}
(ep_1)}{2 N_c Q^2}
\int\limits^1_0 dz\, \phi_\parallel(z) \, .
\eeq
We see that only the longitudinal polarization of the photon contributes.     
The polarization vector for longitudinally polarized photon reads 
\beq{eL}
e_L =\frac{1}{Q}p_1+\frac{Q}{s}p_2 =\frac{1}{Q}p + 2\frac{Q}{s}p_2\,, \quad
(e_L p) = 0\,, \quad e_L^2=1\,.
\eeq
Due to the $U(1)$ gauge invariance, one can omit the first term in the above 
expression and use the simpler vector
\beq{eLnew}
e_L\to e_L =2\frac{Q}{s}p_2\,.
\eeq

The consideration of the other diagrams of Fig.~2 goes along similar lines. 
Here we cite the final result for the hard-scattering amplitude entering the
LLA impact factor
\beq{LO}
T_H^{(0)}(z,\alpha, s_0,\mu_F ,\mu_R)= \alpha_S\frac{ \alpha}{\alpha +z\bar z} 
\ .
\eeq
This result coincides with what found in Ref.~\cite{IG}. Note that method used 
here may be directly applied to the calculation of the impact factors describing 
meson production at nonzero momentum transfer $\vec \Delta$ and for transverse 
polarization. Some results for the corresponding LLA impact factors may be found in
\cite{Ivanov:1998gk,Ivanov:2000uq}.


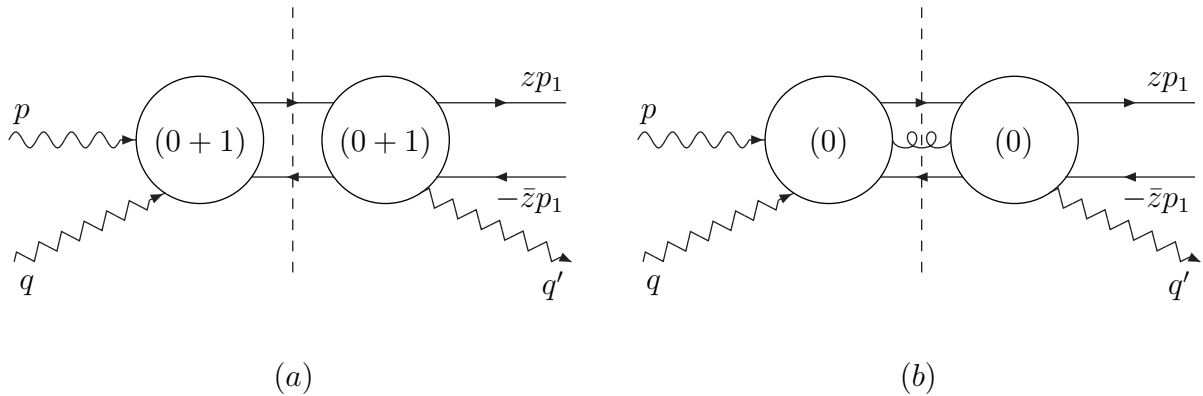
\begin{figure}[tb]
\begin{center}
\begin{picture}(430,120)(0,-20)

\Photon(-12,40)(30,40){3}{4}
\ArrowLine(30,40)(36,40) 
\ArrowLine(79,54)(111,54)
\ArrowLine(111,26)(79,26)
\GCirc(60,40){24}{1}
\GCirc(130,40){24}{1}
\ZigZag(-10,-6)(41,17.264){3}{6}
\ArrowLine(41,17.264)(47,20) 
\ZigZag(143,20)(196,-4.176){3}{6}
\ArrowLine(196,-4.176)(200,-6)
\DashLine(95,90)(95,-10){4}
\ArrowLine(149,54)(198,54)
\ArrowLine(198,26)(149,26)
\Text(-10,50)[l]{$p$}
\Text(-8,-15)[l]{$q$}
\Text(42.7,39)[l]{$(0+1)$}
\Text(112.7,39)[l]{$(0+1)$}
\Text(198,-15)[r]{$q^{\prime}$}
\Text(198,62)[r]{$zp_1$} 
\Text(198,17)[r]{$-{\bar z}p_1$} 

\Text(88,-50)[l]{$(a)$}



\Photon(225,40)(267,40){3}{4}
\ArrowLine(267,40)(273,40) 
\ArrowLine(316,54)(348,54)
\ArrowLine(348,26)(316,26)
\GCirc(297,40){24}{1}
\GCirc(367,40){24}{1}
\Gluon(343,40)(321,40){3}{2}
\ZigZag(227,-6)(278,17.264){3}{6}
\ArrowLine(278,17.264)(284,20) 
\ZigZag(380,20)(433,-4.176){3}{6}
\ArrowLine(433,-4.176)(437,-6)
\DashLine(332,90)(332,-10){4}
\ArrowLine(386,54)(435,54)
\ArrowLine(435,26)(386,26)
\Text(227,50)[l]{$p$}
\Text(229,-15)[l]{$q$}
\Text(290.5,39)[l]{$(0)$}
\Text(360.5,39)[l]{$(0)$}
\Text(435,-15)[r]{$q^{\prime}$}
\Text(435,62)[r]{$zp_1$} 
\Text(435,17)[r]{$-{\bar z}p_1$} 

\Text(325,-50)[l]{$(b)$}

\end{picture}
\end{center}
\vspace*{1cm}
\caption[]{The contributions to the impact factor from two-particle (a) and 
three-particle (b) intermediate states.}
\end{figure}


We see that
due to collinear factorization, which effectively puts some fermion lines 
on the mass-shell, the complexity of the intermediate state contributing to 
the impact factor is reduced in comparison to the case of the virtual photon 
impact factor $\Phi_{\gamma^*\to \gamma^*}$. Actually we have a one-particle state 
instead of a two-particle one in the LLA, and as we will shortly see 
not more than two-particle   
intermediate states in the NLA, whereas a three-particle state contributes 
in the NLA to the $\Phi_{\gamma^*\to \gamma^*}$ impact factor. Due to this, 
the calculation of $\Phi_{\gamma^*\to V}$ is much simpler than that
of $\Phi_{\gamma^*\to \gamma^*}$. This allows to obtain closed analytical
expression for the $\Phi_{\gamma^*\to V}$ impact factor in the NLA, as we
show in the following Section.

\section{Impact factor in NLA}

The impact factor is defined as an integral over the intermediate states
produced in the $\gamma^*$-Reggeon interaction, see Eq.~(\ref{13}).
In the NLA both two particle quark-antiquark $(q \bar q)$ and three-particle
quark-antiquark-gluon $(q\bar q g)$ intermediate states contribute.
They are shown in Fig.~3$(a)$ and Fig.~3$(b)$, respectively.
To calculate the impact factor in the NLA one has to know the  
$(q \bar q)$ production vertices with NLA accuracy and the $(q\bar q g)$ ones 
at the Born level.   

In the previous Section we demonstrated in detail how the impact factor 
factorizes into a convolution of the distribution amplitude $\phi_\parallel 
(z)$ and the hard-scattering amplitude $T_H$, where the latter involves the 
$q \bar q$-pair production on the mass-shell. One can formulate a practical 
rule for the calculation of the meson impact factor. It can be obtained    
considering the impact factor describing the free $q\bar q$ production, 
$\Phi_{\gamma^*\to q\bar q}$, and replacing the quark spinors as follows: 
\beq{int7}
v^i_\alpha (p_1\bar z) \bar u^j_\beta (p_1z)\to \frac{\delta^{ij}}{N_c}
\frac{f_V}{4}
\left( \not\! p_1\right)_{\alpha\beta}\phi_\parallel (z) dz \, .
\eeq


\begin{figure}[tb]
\begin{center}
\begin{picture}(400,320)(-20,-200)

\Photon(-10,86)(37,57.8){3}{4}
\ArrowLine(37,57.8)(41,55.4) 
\GCirc(60,40){24}{1}
\ZigZag(-10,-6)(37,22.2){3}{6}
\ArrowLine(37,22.2)(41,24.6)
\ArrowLine(77,56)(110,56)
\ArrowLine(110,24)(77,24)
\Text(53,39)[l]{$(0)$}
\Text(-8,96)[l]{$p$}
\Text(-8,-15)[l]{$q$}

\Text(130,38)[l]{$=$}

\Photon(150,86)(197,57.8){3}{4}
\ArrowLine(197,57.8)(201,55.4) 
\CArc(220,40)(24,90,270)
\ZigZag(150,-6)(197,22.2){3}{6}
\ArrowLine(197,22.2)(201,24.6) 
\ArrowLine(220,64)(240,64)
\ArrowLine(240,16)(220,16)
\Text(152,96)[l]{$p$}
\Text(152,-15)[l]{$q$}

\Text(255,38)[l]{$+$}

\Photon(280,86)(327,57.8){3}{4}
\ArrowLine(327,57.8)(331,55.4) 
\CArc(350,40)(24,90,270)
\ZigZag(280,-6)(327,22.2){3}{6}
\ArrowLine(327,22.2)(331,24.6) 
\ArrowLine(370,64)(350,64)
\ArrowLine(350,16)(370,16)
\Text(282,96)[l]{$p$}
\Text(282,-15)[l]{$q$}

\Text(175,-40)[l]{$(a)$}

\ArrowLine(-10,-90)(22,-90)
\ArrowLine(22,-122)(-10,-122)
\GCirc(40,-106){24}{1}
\ArrowLine(58,-90)(90,-90)
\ArrowLine(90,-122)(58,-122)
\ZigZag(40,-130)(40,-169){3}{6}
\ArrowLine(40,-170)(40,-173) 
\Text(33,-107)[l]{$(0)$}
\Text(90,-84)[r]{$zp_1$} 
\Text(90,-130)[r]{$-{\bar z}p_1$}
\Text(45,-168)[l]{$q^{\prime}$}

\Text(115,-108)[l]{$=$}
 
\ArrowLine(150,-90)(250,-90)
\ArrowLine(200,-122)(150,-122)
\ArrowLine(250,-122)(200,-122)
\ZigZag(200,-122)(200,-162){3}{6}
\ArrowLine(200,-163)(200,-166)
\Text(250,-84)[r]{$zp_1$} 
\Text(250,-130)[r]{$-{\bar z}p_1$}
\Text(205,-160)[l]{$q^{\prime}$}

\Text(265,-108)[l]{$+$}

\ArrowLine(390,-90)(290,-90)
\ArrowLine(290,-122)(340,-122)
\ArrowLine(340,-122)(390,-122)
\ZigZag(340,-122)(340,-162){3}{6}
\ArrowLine(340,-163)(340,-166)
\Text(390,-84)[r]{$-{\bar z}p_1$} 
\Text(390,-130)[r]{$zp_1$}
\Text(345,-160)[l]{$q^{\prime}$}

\Text(175,-185)[l]{$(b)$}

\end{picture}
\end{center}
\caption[]{The lowest order Feynman diagrams describing Reggeon-photon (a) 
and Reggeon-meson (b) vertices.}
\end{figure}
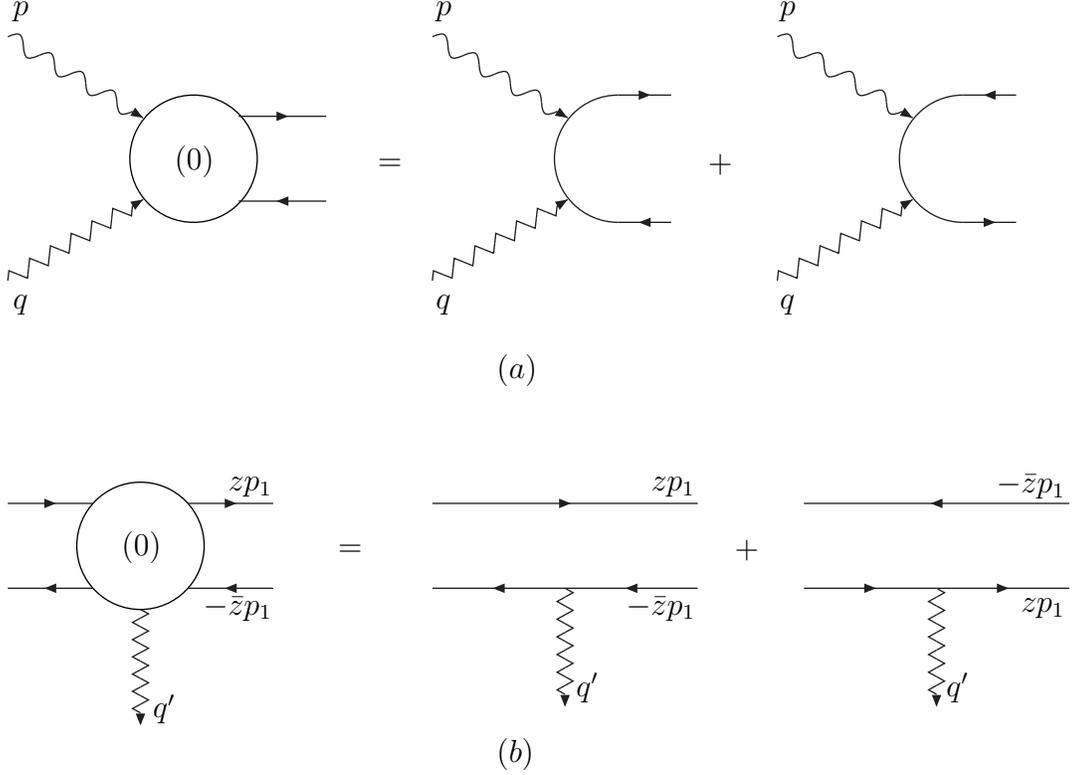


The Born Reggeon-photon vertex is given by the two diagrams in Fig.~4$(a)$
(see~\cite{IG,FIK})
\beq{int8}
\Gamma^{(0)}_{\gamma_L^*q\bar q}=2e_qgQ(t^c)_{ij}z\bar z\left(
\frac{1}{\vec q^{\:2}_{1}+Q^2z\bar z}-\frac{1}{\vec q^{\:2}_{2}+Q^2z\bar z}
\right)
\bar u^i_\alpha(q_1)\left(\frac{\not\! p_2}{s}\right)_{\alpha\beta}
v^j_\beta(q_2) \, ,
\eeq
where $q_1$ and $q_2$ denote the momenta of the quark and of the antiquark.
The contributions to the Reggeon-meson vertex at the leading order are shown
in Fig.~4$(b)$. Due to collinear factorization, quark lines labeled by 
$p_1z$ and $p_1\bar z$ should be included in the definition of the meson    
distribution amplitude $\phi_\parallel (z)$. Therefore, for calculating $T_H$ 
one can use the following expression for the LO Reggeon-meson vertex, 
\[
\left(\Gamma^{(0)}_{V_Lq\bar q}\right)^*=
\frac{gf_V}{4 N_c} 
\, (t^{c^\prime})_{ji}\, \phi_\parallel (z)\, dz 
\]
\beq{int9}
\left[
\bar v^j_\beta(q^\prime+p_1\bar z)
\left(\frac{\not\! p_2}{s}\not\! p_1\right)_{\beta\alpha}
+
\left(\not\! p_1\frac{\not\! p_2}{s}\right)_{\beta
\alpha}\!\!\!u^i_\alpha(p_1z+q^\prime)
\right] \, , 
\eeq
where the first and the second terms represent the contributions of the first 
and the second diagrams of Fig.~4$(b)$, respectively.
They have been drawn as disconnected diagrams, where the upper quark 
(antiquark) line is shown to indicate that, in the convolution of this vertex 
with the Reggeon-photon one, the corresponding quark (antiquark) spinor in the 
expression of the Reggeon-photon vertex~(\ref{int8}) has to be amputated    
because this quark (antiquark) line is a part of $\phi_\parallel (z)$. 
Therefore the two-particle intermediate state $(q\bar q)$ is effectively    
reduced to a one-particle antiquark (quark) state to be cut according to the 
definition of the impact factor. Performing the convolution of the two Born 
vertices given by Eqs.~(\ref{int8}) and (\ref{int9}) one gets again the 
result~(\ref{LO}).


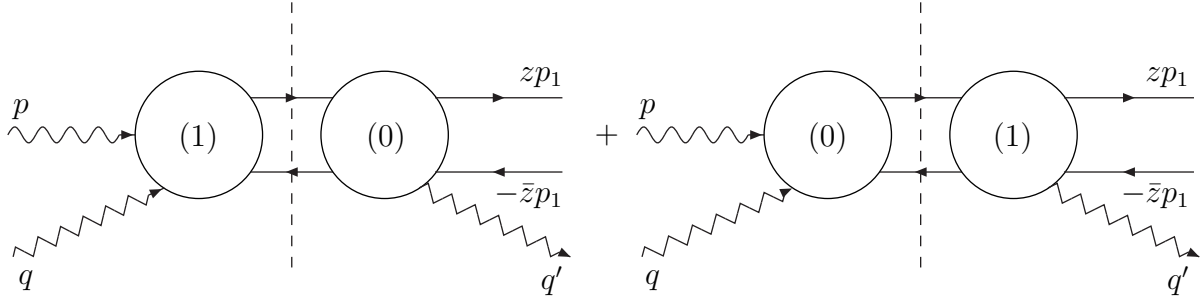
\begin{figure}[tb]
\begin{center}
\begin{picture}(430,140)(0,-40)

\Photon(-12,40)(30,40){3}{4}
\ArrowLine(30,40)(36,40) 
\ArrowLine(79,54)(111,54)
\ArrowLine(111,26)(79,26)
\GCirc(60,40){24}{1}
\GCirc(130,40){24}{1}
\ZigZag(-10,-6)(41,17.264){3}{6}
\ArrowLine(41,17.264)(47,20) 
\ZigZag(143,20)(196,-4.176){3}{6}
\ArrowLine(196,-4.176)(200,-6)
\DashLine(95,90)(95,-10){4}
\ArrowLine(149,54)(197,54)
\ArrowLine(197,26)(149,26)
\Text(-10,50)[l]{$p$}
\Text(-8,-15)[l]{$q$}
\Text(52.5,39)[l]{$(1)$}
\Text(123.5,39)[l]{$(0)$}
\Text(198,-15)[r]{$q^{\prime}$}
\Text(198,62)[r]{$zp_1$} 
\Text(198,17)[r]{$-{\bar z}p_1$}




\Text(210,40)[l]{$+$}

\Photon(225,40)(267,40){3}{4}
\ArrowLine(267,40)(273,40) 
\ArrowLine(316,54)(348,54)
\ArrowLine(348,26)(316,26)
\GCirc(297,40){24}{1}
\GCirc(367,40){24}{1}
\ZigZag(227,-6)(281,18.632){3}{6}
\ArrowLine(281,18.632)(282,19.088)
\ZigZag(380,20)(433,-4.176){3}{6}
\ArrowLine(433,-4.176)(437,-6)
\DashLine(332,90)(332,-10){4}
\ArrowLine(386,54)(435,54)
\ArrowLine(435,26)(386,26)
\Text(227,50)[l]{$p$}
\Text(229,-15)[l]{$q$}
\Text(290.5,39)[l]{$(0)$}
\Text(360.5,39)[l]{$(1)$}
\Text(435,-15)[r]{$q^{\prime}$}
\Text(435,62)[r]{$zp_1$} 
\Text(435,17)[r]{$-{\bar z}p_1$} 


\end{picture}
\end{center}
\caption[]{The contribution of the quark-antiquark intermediate state 
at the next-to-leading order (NLO).}
\end{figure}

The radiation correction to the impact factor may be divided into two parts,
\beq{int10}
T_{H}^{(1)}=T^{(q\bar q)}+T^{(q\bar q g)} \, .
\eeq
Here the first term represents the contribution of the $(q\bar q)$ intermediate
state shown in Fig.~5, when one of the Reggeon effective vertices is taken at 
one-loop order. The second term in Eq.~(\ref{int10}) is the $(q\bar q g)$ 
contribution, see Fig.~3(b), which appears first only in the NLA. 
In that case, it is enough to take the corresponding Reggeon interactions 
at the Born level. We proceed now to the evaluation of the $(q\bar q)$ 
contribution.


\subsection{Quark-antiquark intermediate state}

According to Fig.~5, $T^{(q\bar q)}$ is given by the sum of two terms which 
correspond to the corrections either to the Reggeon-photon or to the 
Reggeon-meson interaction:
\beq{q1}
T^{(q\bar q)}=T^{(q\bar q)}_{\gamma^*}+T^{(q\bar q)}_V \, .
\eeq

The NLA contribution to the Reggeon-meson effective vertex is shown in Fig.~6 
and it splits naturally into the sum of the three different contributions:
\beq{q2}
T^{(q\bar q)}_V=T_V^{(1\bar q)}+T_V^{(1q)}+T_V^{(1c)}\, .
\eeq   
Here the first two terms represent the contributions of disconnected
diagrams which are expressed in terms of NLA effective Reggeon-antiquark and 
Reggeon-quark vertices. They are similar to the disconnected diagrams of
Fig.~4$(b)$ which appear at the Born level.  
In the case of $T_V^{(1\bar q)}$, the two-particle $(q\bar q)$ intermediate
state reduces effectively, due to collinear factorization, to a one-particle
antiquark state, whereas in the case of $T_V^{(1q)}$ it reduces to a 
one-particle quark intermediate state.  
The last term in Eq.~(\ref{q2}) stands for the contribution 
to $T^{(q\bar q)}_V$ from the connected part of the NLA Reggeon-meson
effective vertex, which is shown in  Fig.~7.  The calculation of $T_V^{(1c)}$
involves the integration over the two-particle $(q\bar q)$ state.            


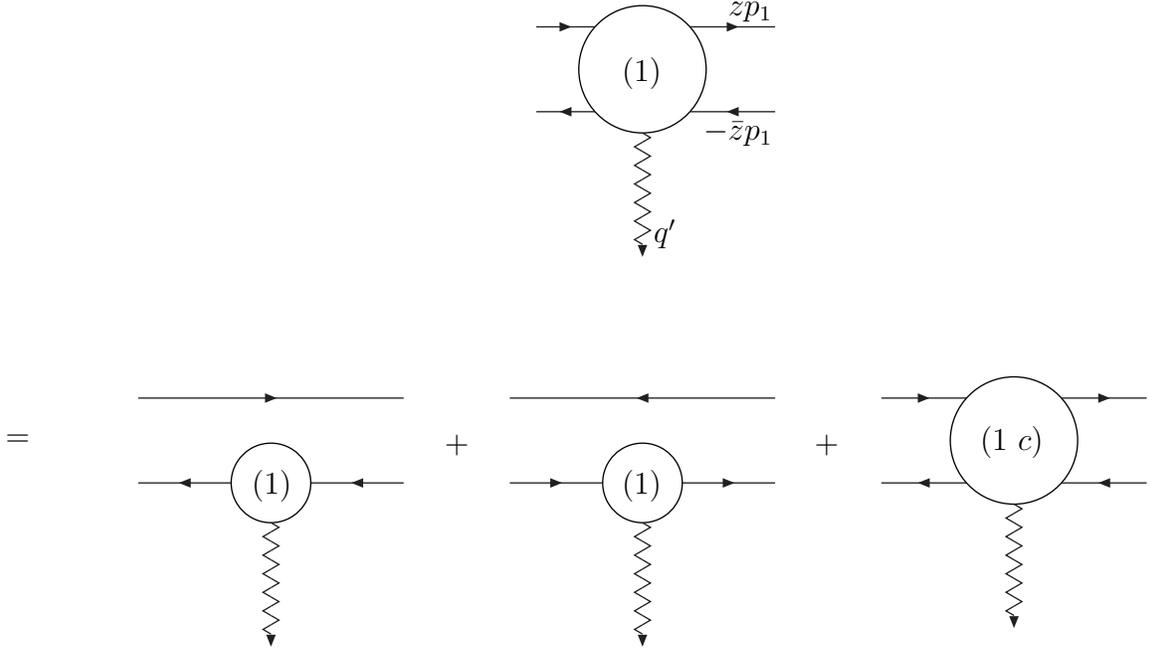
\begin{figure}[tb]
\begin{center}
\begin{picture}(400,240)(-30,-200)

\ArrowLine(160,90)(182,90)
\ArrowLine(182,58)(160,58)
\GCirc(200,74){24}{1}
\ArrowLine(218,90)(250,90)
\ArrowLine(250,58)(218,58)
\ZigZag(200,50)(200,8){3}{6}
\ArrowLine(200,7)(200,4) 
\Text(193,73)[l]{$(1)$}
\Text(250,96)[r]{$zp_1$} 
\Text(250,50)[r]{$-{\bar z}p_1$}
\Text(205,12)[l]{$q^{\prime}$}

\Text(-40,-66)[l]{$=$}

\ArrowLine(10,-50)(110,-50)
\ArrowLine(45,-82)(10,-82)
\ArrowLine(110,-82)(75,-82)
\GCirc(60,-82){15}{1}
\ZigZag(60,-97)(60,-139){3}{6}
\ArrowLine(60,-140)(60,-143)
\Text(53,-83.5)[l]{$(1)$}

\Text(126,-68)[l]{$+$}

\ArrowLine(250,-50)(150,-50)
\ArrowLine(150,-82)(185,-82)
\ArrowLine(215,-82)(250,-82)
\GCirc(200,-82){15}{1}
 \ZigZag(200,-97)(200,-139){3}{6}
\ArrowLine(200,-140)(200,-143)
\Text(193,-83.5)[l]{$(1)$}

\Text(266,-68)[l]{$+$}

\ArrowLine(290,-50)(322,-50)
\ArrowLine(322,-82)(290,-82)
\GCirc(340,-66){24}{1}
\ArrowLine(358,-50)(390,-50)
\ArrowLine(390,-82)(358,-82)
\ZigZag(340,-90)(340,-132){3}{6}
\ArrowLine(340,-133)(340,-136)
\Text(328.5,-67)[l]{$(1~c)$}

\end{picture}
\end{center}
\vspace*{-1.8cm}
\caption[]{The separation of the Reggeon-meson NLO vertex into 
the connected $(1c)$ and the disconnected $(1)$ parts.}
\end{figure}


The virtual photon-Reggeon effective vertices were studied in the NLA in    
\cite{FIK}. Considering  $T^{(q\bar q)}_{\gamma^*}$, we found it convenient 
to distinguish two contributions denoted $T^{(1a)}_{\gamma^*}$ 
and $T^{(1b)}_{\gamma^*}$:
\beq{q3}
T^{(q\bar q)}_{\gamma^*}=T^{(1a)}_{\gamma^*}+T^{(1b)}_{\gamma^*}\, .
\eeq
According to Fig.~8, $T^{(1b)}_{\gamma^*}$ originates from the
particular contribution to the photon-Reggeon vertex given by the 
sum of two box diagrams shown in Fig.~8$(b)$; $T^{(1a)}_{\gamma^*}$ stands 
for the contribution to $T^{(q\bar q)}_{\gamma^*}$ which comes from 
all other contributions to the NLA photon-Reggeon vertex except those two 
box diagrams (for more details see~\cite{FIK}). The reason for such
separation is that we will combine and consider together the connected
contribution to $T_V^{(q\bar q)}$ and the contribution to 
$T^{(q\bar q)}_{\gamma^*}$ coming from the box diagrams. This allows 
to perform the cancellation of some terms in the sum of $T_V^{(1c)}$
and $T^{(1b)}_{\gamma^*}$ in intermediate steps of the calculation. 

Thus, the contribution of the $(q\bar q)$ intermediate state to the 
NLA impact factor may be represented as the sum of three terms 
\beq{q5}
T^{(q\bar q)}=T_1+T_2+T_3 \, ,
\eeq
where 
\beq{q4}
T_1=T^{(1a)}_{\gamma^*}\, , \;\;\;
T_2=T_V^{(1c)}+T^{(1b)}_{\gamma^*}\, , \;\;\;
T_3=T_V^{(1\bar q)}+T_V^{(1q)} \, .
\eeq


\begin{figure}[tb]
\begin{center}
\begin{picture}(400,240)(-25,-190)

\ArrowLine(-10,90)(22,90)
\ArrowLine(22,58)(-10,58)
\GCirc(40,74){24}{1}
\ArrowLine(58,90)(90,90)
\ArrowLine(90,58)(58,58)
\ZigZag(40,50)(40,8){3}{6}
\ArrowLine(40,7)(40,4) 
\Text(28.5,73)[l]{$(1~c)$}
\Text(90,96)[r]{$zp_1$} 
\Text(90,50)[r]{$-{\bar z}p_1$}
\Text(45,12)[l]{$q^{\prime}$}

\Text(120,73)[l]{$=$}

\ArrowLine(150,90)(250,90)
\ArrowLine(250,58)(150,58)
\ZigZag(215,58)(215,16){3}{6}
\ArrowLine(215,15)(215,12)
\Gluon(185,58)(185,90){3}{4}

\Text(194,0)[l]{$(a)$}

\Text(266,72)[l]{$+$}

\ArrowLine(390,90)(290,90)
\ArrowLine(290,58)(390,58)
\ZigZag(355,58)(355,16){3}{6}
\ArrowLine(355,15)(355,12)
\Gluon(325,58)(325,90){3}{4}

\Text(334,0)[l]{$(b)$}

\Text(406,72)[l]{$+$}

\ArrowLine(10,-50)(110,-50)
\ArrowLine(110,-82)(10,-82)
\ZigZag(45,-82)(45,-124){3}{6}
\ArrowLine(45,-125)(45,-128)
\Gluon(75,-82)(75,-50){3}{4}

\Text(54,-140)[l]{$(c)$}

\Text(126,-68)[l]{$+$}

\ArrowLine(250,-50)(150,-50)
\ArrowLine(150,-82)(250,-82)
\ZigZag(185,-82)(185,-124){3}{6}
\ArrowLine(185,-125)(185,-128)
\Gluon(215,-82)(215,-50){3}{4}

\Text(194,-140)[l]{$(d)$}

\Text(266,-68)[l]{$+$}

\ArrowLine(290,-50)(390,-50)
\ArrowLine(390,-82)(290,-82)
\Gluon(340,-105)(320,-50){3}{6}
\Gluon(340,-105)(370,-82){3}{4}
\ZigZag(340,-106)(340,-124){3}{3}
\ArrowLine(340,-125)(340,-128)
 
\Text(334,-140)[l]{$(e)$}

\end{picture}
\end{center}
\vspace*{-1.8cm}
\caption[]{The Feynman diagrams for the connected Reggeon-meson vertex.}
\end{figure}
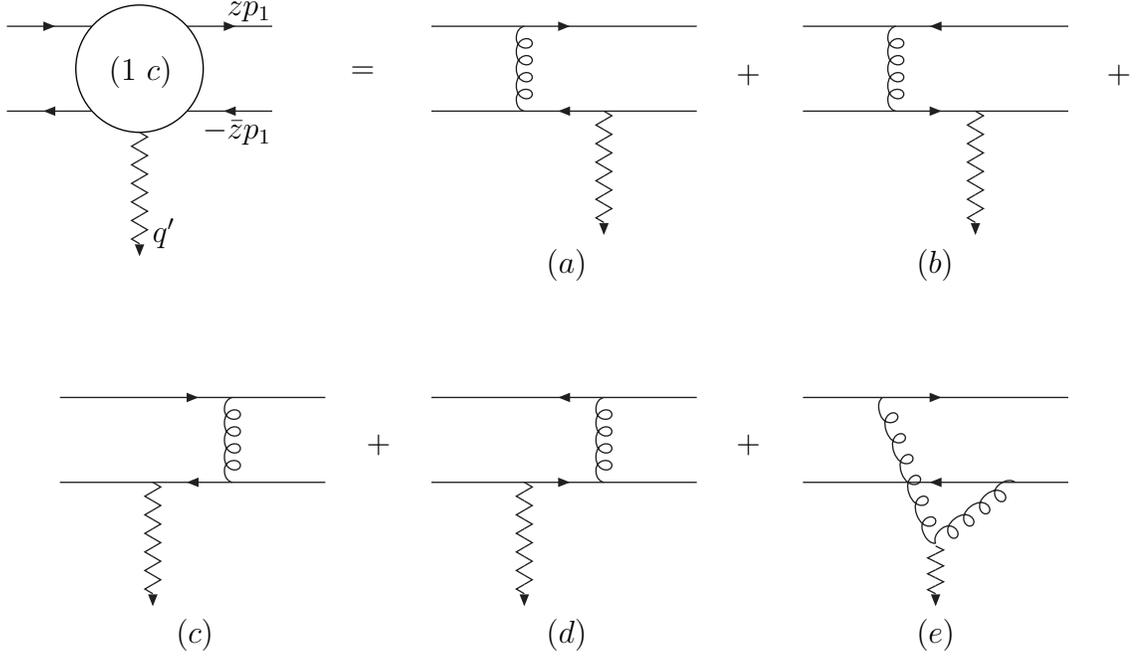

Using the results of~\cite{FIK} and calculating the corresponding 
integrals in our kinematics, we find the following result for $T_1$ 
\beq{T1}
T_1=\frac{g^2\Gamma [1-\varepsilon]
\left(Q^{2}/\mu^2\right)^\varepsilon}{(4\pi)^{2+\varepsilon}}
T_H^{(0)}\left[\, \tau_1(z) + \tau_1(\bar z) \, \right] \, ,
\eeq
where\footnote{Here and in the expressions for $\tau_2(z)$, ..., 
$\tau_5(z)$ which will appear below we used the $z\leftrightarrow \bar z$ symmetry
in order to simplify the results.}
\[
\tau_1(z)=-\frac{\beta_0}{4\,\varepsilon} + N_c\left[-\frac{1}{2\,
\varepsilon^2} +  \frac{3 - 2\ln \alpha - 4\ln z}{4\,\varepsilon} \right] + 
\frac{1}{N_c \varepsilon}\left[-\frac{1}{4} + \frac{\alpha +z\bar z}
{2\,\left( \alpha - z^2 \right)} - \frac{\alpha\,(z+\alpha\bar z)}
{\alpha^2-  z^2 } \right.
\]
\[
\left.
- \frac{z^2} {2\left( \alpha - z^2 \right)^2 }\ln \alpha 
+\frac{\alpha(\alpha +z\bar z)}{2(\alpha -z)^2}\ln\left(\frac{\alpha}{z}\right)
+ \left( \frac{\alpha\,z^3}{2\,\bar z {\left( \alpha +z \right) }^2} 
- \frac{z^2}{2 {\left( \alpha - z^2\right) }^2}\right) 
\ln \left(\frac{\alpha+z\bar z}{\alpha z}\right)\right]  
\]
\[
+\, n_f\left[-\frac{5}{18} + \frac{\ln \alpha}{6} \right]
+ \frac{1}{N_c}\left[1 - \frac{5\,z} {4\,\left( \alpha - z^2 \right) } +    
\frac{5\,z\,\left( \alpha + z\bar z \right) }{2\,\left(\alpha^2 - z^2\right)}
- \left( \frac{2-z}{4\bar z} + \frac{\left(2 - z\right)z }{2\,{\left( \alpha 
- z\right) }} \right.\right.
\]
\[
\left. + \frac{5\,z^4} {4\,\bar z \,{\left( \alpha + z\right) }^2} 
-\frac{z^2\,\left(2+ 3\,z \right) }{4\,\bar z
\left( \alpha + z \right) }\right) \ln \alpha + \frac{1-2z\bar z}{8\alpha}
\ln\left(\frac{\alpha+z\bar z}{z \bar z}\right) 
\]
\[
+ \left(\frac{\alpha\,(\alpha +z\bar z)}{4\,{\left( \alpha - z\right) }^2} - 
\frac{\alpha\, z^3}{4\,\bar z \left( \alpha + z\right)^2 } \right) 
{\ln^2 \alpha} - \frac{\alpha (\alpha+z\bar z)}{4(\alpha -z)^2}\ln^2 z
\]
\[
+ \left(\frac{1 - 2\,z}{4\,\bar z} + \frac{7\,z - z^2} {4\,\left( \alpha - z
\right)}+\frac{5\,\left( 2\,z^2 - z^3\right)}{4\,{\left(\alpha - z\right) }^2}
\right) \,\ln \left(\frac{z}{\alpha}\right)
\]
\[
+\left(\frac{5\, z^4}{4\,\bar z\left( \alpha + z \right)^2}-\frac{4\,z^2 + z^3}
{4\,\bar z\,\left(\alpha + z\right)}+ \frac{6\,z^2-4\,z^3-\alpha+4\,z\,\alpha}
{4\,{\left( \alpha - z^2\right) }^2}\right) \,\ln \left(\frac{\alpha+z\bar z}
{z}\right)   
\]
\[
\left. + \left(\frac{\alpha\,z^3}{4\,\bar z\,{\left( \alpha + z\right) }^2}
- \frac{z^2}{4\, {\left(\alpha- z^2 \right) }^2}\right) 
{\ln^2 \left(\frac{\alpha+z\bar z}{z}\right)}\right]
\]
\[
+N_c\left[\frac{1}{9} + \frac{\alpha(\alpha+z\bar z)}{4\,\left(\alpha -z\right)
\,\left( \alpha - z^2 \right)}- \frac{3}{4}\ln^2\alpha-\frac{1}{2}
\ln^2\left(\frac{\alpha+z\bar z}{z}\right) \right. 
\]
\[
+ \left(\frac{z^3}{4\,\bar z\,{\left( \alpha - z^2 \right) }^2} + \frac{z}
{2\,\left( \alpha - z^2 \right)} - \frac{z^2}{4\,\bar z \,\left( \alpha + z
\right) } \right) \ln \left(\frac{\alpha +z\bar z}{z}\right) 
- \left(\frac{1}{6} + \frac{\alpha+z\bar z}{4\bar z \left( \alpha + z\right)} 
\right)\ln\,\alpha
\]
\[
+ \frac{\alpha^2-2\,\bar z\,z^2 - \alpha\,z(1+z)}{4\,\bar z \left(\alpha - z
\right)^2}\ln\left(\frac{\alpha}{z}\right)-\ln\left(\frac{\alpha+z}{z^2}\right)
\ln\left(\frac{\alpha+z\bar z}{\alpha z}\right)
\]
\beq{tau1}
\left.
- {\rm Li}_2\left(-\frac{z}{\alpha}\right) 
+ {\rm Li}_2\left(-\frac{z\bar z}{\alpha}\right) 
- {\rm Li}_2\left(z - \frac{\alpha}{z}\right)      
- {\rm Li}_2\left(- \frac{\alpha + z \bar z }{z^2}\right) \right]
\eeq
and 
\beq{li2}
{\rm Li}_2(z)=-\int\limits^z_0\frac{dt}{t}\ln(1-t) \ .
\eeq

Our result for $T_2$ is
\beq{T2}
T_2=\frac{g^2\Gamma [1-\varepsilon]
\left(Q^{2}/\mu^2\right)^\varepsilon}{(4\pi)^{2+\varepsilon}}
T_H^{(0)}\left[\, \tau_2(z) + \tau_2(\bar z) \, \right] \, ,
\eeq
where
\[
\tau_2(z)=\frac{\alpha+z\bar z}{\alpha} \int\frac{d^{2+2\varepsilon}\vec
l\left(Q^2\right)^{1-\varepsilon}}{\pi^{1+\varepsilon}
\Gamma[1-\varepsilon]}\int\limits^1_0 du
\frac{u\bar u}{z\bar z}
\]
\[
\times \left[
\left(
\frac{1-\varepsilon}{4}C_F+\frac{1}{4N_c}
\right)
\left(
\frac{1}{\left(\vec l-\vec q\right)^2}-\frac{1}{\vec l^{\:2}}
\right)
\frac{1}{\vec l^{\:2}+Q^2u\bar u }
\right.
\]
\[
-\frac{1}{2N_c}
\left(
\frac{\bar z(u-z)\left[(1+\varepsilon )\bar z \vec l^{\:2}
-\left(1+(1+\varepsilon)(u-z)\right)(\vec l\vec q) +u \vec q^{\:2}\right]}
{\vec l^{\:2}\left[\left(\vec l-\vec q\right)^2+Q^2u\bar u\right]
\left(\vec l\bar z-(u-z)\vec q\right)^2}
\right.
\]
\[
+ \frac{\bar uz^3\bar z^2
\left[(1+\varepsilon )\bar z \vec l^{\:2}
-\left(1-(1+\varepsilon)z\bar u\right)(\vec l\vec q) +uz \vec q^{\:2}\right]}
{\vec l^{\:2}\left[\bar z
\left(\vec l-\bar u \vec q\right)^2+u\bar u\left(
\vec q^{\:2}+Q^2z\bar z\right)\right]\left(\vec l\bar z+z\bar u\vec q\right)^2}
\]
\beq{tau20}
-\left.\left.\frac{u z^2
\left[(1+\varepsilon )z \vec l^{\:2}
-\left(1+(1+\varepsilon)zu\right)(\vec l\vec q) +(1-\bar u z) 
\vec q^{\:2}\right]}
{\vec l^{\:2}\left[
\vec l^{\:2}+Q^2u\bar u z\right]\left(\vec l- u\vec q\right)^2}
\right)\right] \, . 
\eeq
Note that the box diagrams in 
Fig.~8(b) and the diagrams Fig.~7(a) and Fig.~7(b) contain imaginary
parts. But combined together in $T_2$ these imaginary parts cancel.
We made this cancellation explicit proceeding in the following way. 
In the loop integrals for box diagrams we split the loop integration
into the integral over the transverse part of the loop momentum and the
integral over the two Sudakov variables. Then, we  performed  
the integration over one of the Sudakov variables. After that,
we combined the results for box diagrams with all other contributions to $T_2$ 
and using some algebraic transformations obtained the representation (\ref{tau20}). 


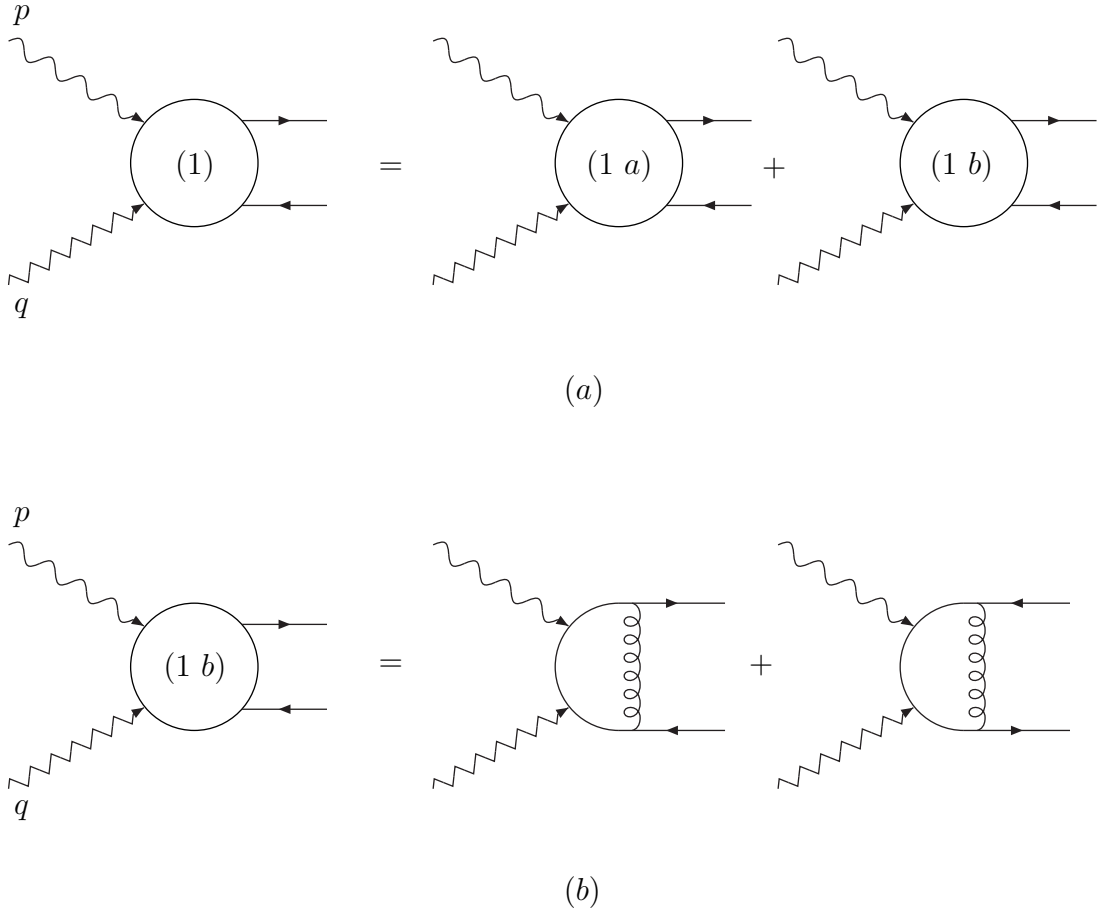
\begin{figure}[tb]
\begin{center}
\begin{picture}(450,300)(-20,-270)

\Photon(-10,86)(37,57.8){3}{4}
\ArrowLine(37,57.8)(41,55.4)
\GCirc(60,40){24}{1}
\ZigZag(-10,-6)(37,22.2){3}{6}
\ArrowLine(37,22.2)(41,24.6)
\ArrowLine(78,56)(110,56)
\ArrowLine(110,24)(78,24)
\Text(53,39)[l]{$(1)$}
\Text(-8,96)[l]{$p$}
\Text(-8,-15)[l]{$q$}

\Text(130,38)[l]{$=$}

\Photon(150,86)(197,57.8){3}{4}
\ArrowLine(197,57.8)(201,55.4) 
\GCirc(220,40){24}{1}
\ZigZag(150,-6)(197,22.2){3}{6}
\ArrowLine(197,22.2)(201,24.6)
\ArrowLine(238,56)(270,56)
\ArrowLine(270,24)(238,24)
\Text(208.5,39)[l]{$(1~a)$}

\Text(275,38)[l]{$+$}

\Photon(280,86)(327,57.8){3}{4}
\ArrowLine(327,57.8)(331,55.4) 
\GCirc(350,40){24}{1}
\ZigZag(280,-6)(327,22.2){3}{6}
\ArrowLine(327,22.2)(331,24.6)
\ArrowLine(368,56)(400,56)
\ArrowLine(400,24)(368,24)
\Text(338.5,39)[l]{$(1~b)$}

\Text(200,-46)[l]{$(a)$}

\Photon(-10,-104)(37,-132.2){3}{4}
\ArrowLine(37,-132.2)(41,-134.6) 
\GCirc(60,-150){24}{1}
\ZigZag(-10,-196)(37,-167.8){3}{6}
\ArrowLine(37,-167.8)(41,-165.4)
\ArrowLine(78,-134)(110,-134)
\ArrowLine(110,-166)(78,-166)
\Text(48.5,-151)[l]{$(1~b)$}
\Text(-8,-94)[l]{$p$}
\Text(-8,-205)[l]{$q$}

\Text(130,-150)[l]{$=$}

\Photon(150,-104)(197,-132.2){3}{4}
\ArrowLine(197,-132.2)(201,-134.6) 
\CArc(220,-150)(24,90,270)
\ZigZag(150,-196)(197,-167.8){3}{6}
\ArrowLine(197,-167.8)(201,-165.4) 
\ArrowLine(220,-126)(260,-126)
\ArrowLine(260,-174)(220,-174)
\Gluon(225,-126)(225,-174){3}{6}

\Text(270,-150)[l]{$+$}

\Photon(280,-104)(327,-132.2){3}{4}
\ArrowLine(327,-132.2)(331,-134.6) 
\CArc(350,-150)(24,90,270)
\ZigZag(280,-196)(327,-167.8){3}{6}
\ArrowLine(327,-167.8)(331,-165.4) 
\ArrowLine(390,-126)(350,-126)
\ArrowLine(350,-174)(390,-174)
\Gluon(355,-126)(355,-174){3}{6}

\Text(200,-236)[l]{$(b)$}

\end{picture}
\end{center}
\vspace*{-1.3cm}
\caption[]{$(a)$ The separation of the NLO virtual photon-Reggeon vertex    
into the sum of two contributions $(1a)$ and $(1b)$. $(b)$  
The contribution $(1b)$ is given by two box diagrams.}
\end{figure}


The integration of (\ref{tau20}) gives
\[
\tau_2(z)=\frac{C_F}{\varepsilon}\left[-\frac{\alpha + z\bar z}{4\,c\,z\bar z}
\, \ln \left(\frac{2\,c+1}{2\,c-1 } \right) \right]+\frac{1}{N_c}
\left[\frac{1}{2\,\varepsilon^2} + \frac{1}{\varepsilon}
\left(\frac{z\left(\alpha+z\bar z \right) }{\alpha^2 - z^2} \right.\right.
\]
\[
- \frac{\alpha+z\bar z} {2\left( \alpha - z^2 \right)} 
+ \left(\frac{1}{2}
+ \frac{z^2} {2\, {\left( \alpha - z^2 \right) }^2} \right) \ln \alpha
+ \frac{\alpha\, (\alpha+z\bar z)}{2\, (\alpha-z)^2} 
\ln \left(\frac{z}{\alpha}\right)
\]
\[
+
\left.\left.\left(\frac{z^2}{2\left( \alpha - z^2 \right)^2} 
- \frac{\alpha z^3}{{2\bar z\left( \alpha + z\right)}^2}\right) 
\,\ln \left(\frac{\alpha + z\bar z}{\alpha z}\right)\right)\right]
\]
\[
+C_F \left[\frac{\alpha + z\bar z}{2z\bar z}\left(\frac{1}{\alpha}
\left(2\,c\,\ln \left(\frac{2\,c+1}{ 2\,c - 1}\right)+\ln \alpha \right)    
+ \frac{1}{c}\left(\frac{{\pi }^2}{6} + \ln \alpha\ln 4 - {\ln^2 c} 
\right.\right.\right.
\]
\[
+ 2\ln c \ln \left(\frac{2\,c-1}{4}\right) - {\ln^2 (2\,c+1)} 
+ \frac{1}{2}\ln \left(\frac {2\,c+1}{2\,c-1}\right) + \frac{1}{4}\ln^2 
\left(\frac{2\,c+1}{2\,c-1}\right)          
\]
\[
- \left.\left.\left. 2  {\rm Li}_2\left(\frac{ 2\,c-1}{4\,c}\right) - 
{\rm Li}_2\left(\frac{2}{1 + 2\,c}\right) \right)\right)\right]
+ \frac{1}{N_c} \left[ \frac{\alpha +z\bar z}{4z^2\bar z^2}
\left(  2\,c\, \ln \left(\frac{2\,c+1}{2\,c-1} \right) +\ln\alpha-2\right)
\right.
\]
\[
- \frac{ \alpha^2 + \alpha\,z -\bar z\,z^3}{2z^3\bar z}
\left( \ln \left(\frac{\alpha \, z}{\alpha + z\bar z}\right)
\ln \left(\frac{\alpha + z}{\alpha}\right) + \ln \left(\frac{2\,c+1}
{2\,c-1}\right)\,\ln \left(\frac{2\,\alpha + z\left( 1 + 2\,c \right)}
{2\,\alpha + z\left( 1 - 2\,c \right)}\right.\right)  
\]
\[
+ \frac{\pi^2}{6} + 2 {\rm Li}_2\left(\frac{2\,z}{1-2\,c}\right)
+2{\rm Li}_2\left(\frac{2\,z}{1+2\,c}\right) - {\rm Li}_2\left(-\frac{z}
{\alpha }\right) + {\rm Li}_2\left(-\frac{z^2 }{\alpha + z \bar z }\right)
\]
\[
\left. 
+ {\rm Li}_2\left(z-\frac{\alpha}{z}\right) +\frac{1}{2}\ln^2\left(
\frac{\alpha+z\bar z}{\alpha} \right)\right)  
- \frac{\pi^2}{12}+\frac{3+2z}{2z}
+ \frac{\alpha}{2\,z^2\,{\bar z }^2} - \frac{8\, z-9\,z^2}
{4\bar z\left(\alpha - z \right) } 
\]
\[
+ \frac{5\,z - 4\,z^2 }{4\bar z(\alpha -z^2)} + \frac{5\, z^3}{2(\alpha^2-z^2)}
+ \left(\frac{5\,z^4} {4\, \bar z \, {\left( \alpha + z \right)}^2} 
+ \frac{3\,z - z^2 - \alpha }{2\,\left( \alpha - z \right)} 
- \frac{3\,z^2+3\,z^3}{4\,\bar z \,\left( \alpha + z \right)}\right) \ln\alpha 
\]
\[
+ \left(\frac{1}{4}+\frac{\alpha\,z^3}{4\,\bar z \,{\left(\alpha + z\right)}^2}
\right) {\ln^2 \alpha} 
+ \frac{1-2z\bar z}{8\alpha}\ln \left(\frac{z\bar z}{\alpha+z\bar z}\right)+ 
\frac{1}{4 z\bar z} \ln\, (\alpha +z\bar z)
\]
\[
+ \left( \frac{1} {4} + \frac{3\,z-z^2}{4 \,{\left( \alpha - z \right) }} 
+ \frac{2\,z^2 - z^3} {4\,(\alpha-z)^2} \right)\left(\ln^2 z-\ln^2\alpha\right)
\]
\[
- \left(\frac{1}{2}+\frac{4\,z-7\,z^2 +z^3}{4\,\bar z\,{\left(\alpha-z\right)}}
+ \frac{8\,z^2 - 14\,z^3 + 5\,z^4} {4\,\bar z {\left( \alpha - z \right)}^2\,} 
\right) \ln \left(\frac{z}{\alpha}\right)
\]
\[
+ \left( \frac{5\,z^2+z^3}{4\bar z(\alpha +z)} -\frac{5\,z^4}{4\,\bar z \,
{\left( \alpha + z \right)}^2}  - \frac{5\,z^2-4\,z^3}{4\,\bar z{\left(\alpha 
- z^2 \right) }^2} + \frac{1-5\, z+6\,z^2}{4\,\bar z\left(\alpha - z^2 \right)}
\right) \ln \left(\frac{\alpha +z\bar z}{z}\right) 
\]
\beq{tau2}
+
\left.\left(\frac{1}{4}-\frac{\alpha z^3}{4\,\bar z\,{\left(\alpha+z\right)}^2}
- \frac{\alpha\, \left( \alpha + z \right)}{4\,\bar z z^3 } 
+\frac{z^2}{4(\alpha-z^2)^2}\right)\ln^2\left(\frac{\alpha+z\bar z}{z}\right)
\right]\;,
\eeq
where we used the notation $c=\sqrt{\alpha+1/4}\,\,$.

Finally, using the result for the NLA quark-quark-Reggeon effective 
vertex~\cite{FFQ94} we obtain 
\beq{T3}
T_3=\frac{g^2\Gamma [1-\varepsilon]
\left(Q^{2}/\mu^2\right)^\varepsilon}{(4\pi)^{2+\varepsilon}}
T_H^{(0)}\left[\, \tau_3 (z) + \tau_3(\bar z) \, \right] \, ,
\eeq
where 
\[
\tau_3(z)=-\frac{\beta_0}{4\,\varepsilon}
+C_F\left[-\frac{1}{\varepsilon^2}
+\frac{3}{2\,\varepsilon}-\frac{\ln\alpha+2\ln z}{\varepsilon}\right]
-\frac{\ln z}{N_c\varepsilon}
+ n_f\left[-\frac{5}{18}+\frac{\ln \alpha}{6}\right]
\]
\beq{tau3}
+C_F\left[
\frac{13}{18}+\frac{2\pi^2}{3}-\frac{\ln\alpha}{3}-\frac{1}{2}\ln^2\alpha
-2\ln \alpha\ln z \right]
+\frac{1}{N_c}\left[\frac{85+9\pi^2}{36}-\frac{11\ln\alpha}{12}-\ln\alpha\ln z
\right]\;.
\eeq


\begin{figure}[tb]
\begin{center}
\begin{picture}(450,400)(-20,-320)

\Photon(-44,40)(-2,40){3}{4}
\ArrowLine(-2,40)(4,40) 
\ArrowLine(47,54)(165,54)
\ArrowLine(88,26)(47,26)
\ArrowLine(165,26)(118,26)
\GCirc(28,40){24}{1}
\GCirc(103,26){15}{1}
\ZigZag(-42,-6)(9,17.264){3}{6}
\ArrowLine(9,17.264)(15,20) 
\ZigZag(116,18.507)(164,-4.176){3}{6}
\ArrowLine(164,-4.176)(168,-6)
\DashLine(67.5,90)(67.5,-10){4}
\Gluon(52,40)(90,33.493){3}{4}
\Text(-42,50)[l]{$p$}
\Text(-40,-15)[l]{$q$}
\Text(20.5,39)[l]{$(0)$}
\Text(96,25.5)[l]{$(0)$}
\Text(166,-15)[r]{$q^{\prime}$}
\Text(166,62)[r]{$zp_1$} 
\Text(166,17)[r]{$-{\bar z}p_1$}

\Text(200,40)[l]{$+$}

\Photon(240,40)(282,40){3}{4}
\ArrowLine(282,40)(288,40)
\ArrowLine(449,54)(331,54)
\ArrowLine(331,26)(372,26)
\ArrowLine(402,26)(449,26)
\GCirc(312,40){24}{1}
\GCirc(387,26){15}{1}
\ZigZag(240,-6)(293,17.264){3}{6}
\ArrowLine(293,17.264)(299,20) 
\ZigZag(400,18.507)(448,-4.176){3}{6}
\ArrowLine(448,-4.176)(452,-6)
\DashLine(351.5,90)(351.5,-10){4}
\Gluon(336,40)(374,33.493){3}{4}
\Text(242,50)[l]{$p$}
\Text(244,-15)[l]{$q$}
\Text(304.5,39)[l]{$(0)$}
\Text(380,25.5)[l]{$(0)$}
\Text(450,-15)[r]{$q^{\prime}$}
\Text(450,62)[r]{$-{\bar z}p_1$} 
\Text(450,17)[r]{$zp_1$}

\Text(195,-50)[l]{$(a)$}

\ArrowLine(186,-130)(150,-130)
\ArrowLine(252,-130)(216,-130)
\GCirc(201,-130){15}{1}
\ZigZag(214,-137.493)(262,-160.176){3}{6}
\ArrowLine(262,-160.176)(268,-162) 
\Gluon(150,-116)(188,-122.507){3}{4}
\Text(194.5,-130.5)[l]{$(0)$}
\Text(266,-171)[r]{$q^{\prime}$}
\Text(253,-122)[r]{$-{\bar z}p_1$}

\Text(-50,-240)[l]{$=$}

\ArrowLine(90,-240)(-12,-240)
\Gluon(-11,-226)(23,-240){3}{4}
\ZigZag(55,-240)(103,-262.683){3}{6}
\ArrowLine(103,-262.683)(109,-264.507) 
\Text(91,-232)[r]{$-{\bar z}p_1$}
\Text(107,-273.507)[r]{$q^{\prime}$}

\Text(115,-241)[l]{$+$}

\ArrowLine(252,-240)(150,-240)
\Gluon(195,-226)(227,-240){3}{4}
\ZigZag(175,-240)(223,-262.683){3}{6}
\ArrowLine(223,-262.683)(229,-264.507) 
\Text(253,-232)[r]{$-{\bar z}p_1$}
\Text(227,-273.507)[r]{$q^{\prime}$}

\Text(277,-241)[l]{$+$}

\ArrowLine(362,-225)(312,-200)
\ArrowLine(412,-225)(362,-225)
\Gluon(362,-225)(362,-255){3}{3}
\Gluon(312,-280)(362,-255){3}{4}
\ZigZag(362,-255)(410,-277.683){3}{6}
\ArrowLine(410,-277.683)(416,-279.507) 
\Text(413,-217)[r]{$-{\bar z}p_1$}
\Text(414,-288.507)[r]{$q^{\prime}$}

\Text(195,-323)[l]{$(b)$}

\end{picture}
\end{center}
\vspace*{0.3cm}
\caption[]{(a) The separation of the $(q\bar q g)$ contribution
to the impact factor into the sum of 
of the antiquark-gluon and the quark-gluon cuts. (b) The lowest order
Feynman diagrams for the gluon bremsstrahlung vertex.}
\end{figure}


\subsection{Quark-antiquark-gluon intermediate state}

We proceed now to the evaluation of the contribution to the impact
factor from the $(q\bar q g)$ intermediate
state shown in Fig.~3(b). For this purpose we need the $(q\bar q g)$ 
production vertices at the Born level. Examining the Feynman diagrams for
the corresponding Reggeon-meson vertex, one can easily see that at the Born level
only disconnected diagrams contribute, i.e. in each diagram either the quark or the
antiquark cut line enters directly the meson vertex and therefore is absorbed
into the definition of the meson distribution amplitude. Thus the 
Reggeon-meson vertex can be represented as the sum of two contributions,
\beq{decmesonV}
\Gamma_{V_Lq\bar q g}=\Gamma^q_{V_Lq\bar q g}+\Gamma^{\bar q}_{V_Lq\bar q g} \, .
\eeq
Consequently, one deals here with the contributions of effective two-particle 
intermediate states, the antiquark-gluon one (related to $\Gamma^q_{V_Lq\bar qg}$) 
and the quark-gluon one (related to $\Gamma^{\bar q}_{V_Lq\bar qg}$), 
which are shown in the first and the second diagrams in Fig.~9(a).
Since due to the charge conjugation symmetry these two contributions are  
related to each other, it is enough to evaluate one of them, say the
first diagram of Fig.~9(a), with the effective antiquark-gluon cut. The
result for the diagram with the effective quark-gluon cut is obtained from 
the contribution of the antiquark-gluon one by the replacement $z\to \bar z$. 
Therefore, in what follows we will consider the antiquark-gluon cut.      

According to the first diagram in Fig.~9(a),
to calculate $\Gamma^q_{V_Lq\bar qg}$ one needs to consider the gluon 
bremsstrahlung from the antiquark line, described by the three
diagrams shown in Fig.~9(b). After a simple calculation we obtain
the following result for the corresponding part of Reggeon-meson vertex 
\beq{int99}
\left(\Gamma^{q}_{V_Lq\bar q g}\right)^*=
-\frac{g^2f_V}{4 N_c}
\bar v^j_\beta(q_2)
\left(\left[
2\bar z\left(\vec T \vec e^{\,\, a}\right)+z_3\not\! 
e^{\,a}_\perp\!\!\not\! T\right]
\frac{\not\! p_2}{s}\not\!
p_1\right)_{\beta\alpha}
\, \phi_\parallel (z)\, dz \, .
\eeq
Here $q_2=z_2 p_1 + (\vec q_{2}^{\:2}/z_2) (p_2/s)+ q_{2\perp}$
denotes the momentum of the cut antiquark, the momentum of the cut gluon is
$k=z_3 p_1+ (\vec k^2/z_3) (p_2/s)+ k_\perp$,
\beq{z2z3}
z_2+z_3=1-z\, , \quad\quad\quad  k_\perp+q_{2\perp}=q_\perp \, .
\eeq
The polarization state of the cut gluon is described by the 4-vector
$e^a$, which can be expressed in the gauge $e^a p_2=0$ as
\beq{ea}
e^a=\frac{2(\vec e^{\, a}\vec k)}{z_3\, s}p_2+ e^a_{\perp}\, , \quad
e^a k=0 \, ,
\eeq
where the index $a$ describes the color state of the gluon, 
and the transverse vector $e^a_\perp$ parameterizes two different physical
polarization states of the gluon,  $(e^{\,\,a}_\perp)^2=-(\vec e^{\,\,a})^2$. 
The other transverse vector entering~(\ref{int99}) is 
\beq{TT}
\vec T=\left(\frac{\bar z\vec q_2-z_2\vec q}{(\bar z\vec q_2-z_2\vec q\, )^2}
-\frac{\vec q_2}{\bar z \vec q_2^{\:2}}\right) (t^at^{c^\prime})_{ji}
+\frac{1}{\bar z}\left(\frac{\vec k}{\vec k^2}
+\frac{\vec q_2}{\vec q_2^{\:2}}\right)(t^{c^\prime}t^a)_{ji} \, ,             
\eeq
where the indices $i,j$ describe the color states of the quark and of the antiquark,
respectively. 

The Reggeon-photon $(q\bar q g)$ production vertex was considered 
in \cite{FIK}. Similarly to the Reggeon-meson vertex discussed above,
we rewrite the Reggeon-photon vertex in terms of the transverse 
gluon polarization vector $e^a_\perp$ and of other two transverse vectors
which we denote $\vec T_q$ and $\vec T_{\bar q}$:
\beq{phvertex}
\Gamma_{\gamma^*_Lq\bar q g}=-2\, e_q\, g^2\, Q\, 
\bar u^i_\alpha(z p_1)
\left(\left[
2\left(\vec e^{\,\, a}(\vec T_q\bar z-\vec T_{\bar q } z)\right)+z_3\left(
\not\! T_q+\not\! T_{\bar q} \right)\not\!
e^{\,a}_\perp\right]
\frac{\not\! p_2}{s}\right)_{\alpha\beta}\!\!\!v^j_\beta(q_2) \, ,
\eeq
where
\[
\vec T_q = 
\left(
\frac{\bar z \vec  q_2-z_2\vec q}{(\bar z \vec  q_2-z_2\vec q)^2}
\left(\frac{1}{Q^2}-\frac{1}{ Q^2+\frac{\vec q^{\,2}}{z}+
\frac{\vec q_2^{\,2}}{z_2}+\frac{\vec k^2}{z_3} }\right)
- \frac{\vec q_2}{Q^2+\frac{\vec q_2^{\,2}\bar z}{z_2z_3}}\frac{1}{Q^2z_2z_3}
\right)(t^ct^a)_{ij} 
\]
\beq{Tq}
+
\frac{1}{Q^2 z_2 z_3}
\left(
\frac{\vec k}{ Q^2 + \frac{\vec k^2\bar z}{z_2 z_3} }+
\frac{\vec q_2}{ Q^2 + \frac{\vec q_2^{\,2} \bar z}{z_2 z_3}}
\right)(t^at^{c})_{ij} \, ,
\eeq
and
\[
\vec T_{\bar q} = (t^ct^a)_{ij}
\left(
\frac{ z\vec q_2-\bar z_2\vec q }{ (z\vec q_{2}-\bar z_2\vec q )^2 }
\left(
\frac{1}{ Q^2+\frac{\vec q_2^{\,2}}{z_2\bar z_2} } -
\frac{1}{ Q^2+\frac{\vec q^{\,2}}{z}+
\frac{\vec q_2^{\,2}}{z_2}+\frac{\vec k^2}{z_3} }
\right) +
\frac{\vec q_2}{Q^2 + \frac{\vec q_2^{\,2}}{z_2\bar z_2}}
\frac{1}{Q^2 z z_2\bar z_2} \right.
\]
\beq{Tbarq}
\left.
-\frac{\vec q_2}{Q^2 + \frac{\vec q_2^{\,2}\bar z}{z_2 z_3}}
\frac{\bar z}{Q^2 z z_2 z_3}
\right)
+
\frac{(t^at^{c})_{ij}}{z}
\left(
\frac{\vec k}{\vec k^2}
\left(
\frac{1}{Q^2+\frac{\vec q_2^{\,2}}{z_2\bar z_2}} -
\frac{1}{Q^2}
\right)  \right.
\eeq
\[
\left.
+\frac{\bar z}{Q^2 z_2 z_3}\left(
\frac{\vec k}{Q^2 +
\frac{\vec k^2\bar z}{z_2 z_3}} 
+
\frac{\vec q_2}
{Q^2 + \frac{\vec q_2^{\,2}\bar z}{z_2 z_3}}\right)
 -
\frac{\vec q_2}{Q^2 + \frac{\vec q_2^{\,2}}{z_2\bar  z_2}}
\frac{1}{Q^2 z_2\bar z_2}
\right)\, ,
\]
with $\bar z_2=1-z_2=z+z_3$.

Note that the quantities $\vec T$, $\vec T_q$ and $\vec T_{\bar q}$
and, therefore, the Reggeon vertices $\Gamma_{V_L q\bar q g}$ and 
$\Gamma_{\gamma_L^* q\bar q g}$, vanish when the Reggeon transverse momentum 
$\vec q$ tends to zero.

The convolution\footnote{Beforehand, the quark spinor $\bar u^i_\beta (p_1 z)$ 
has to be amputated from the photon vertex, since it belongs to the definition
of the meson distribution amplitude.} of the Reggeon vertices~(\ref{phvertex})
and~(\ref{int99}), with the subsequent summation over the Dirac and the color
indices of the cut antiquark and gluon, gives the following contribution  
of the diagrams in Fig.~9(a) to the 
impact factor:
\bea{zzzz}
&&
-\frac{e_qg^4f_VQ}{N_c(2\pi)^{3+2\varepsilon}}
\int\limits^{1}_0 \, \phi_\parallel (z)\, dz
 \int\limits^{\bar z}_0
\frac{dz_3}{z_3}d^{2+2\varepsilon}\vec q_{2} \: \theta(s_\Lambda - \kappa)
\nonumber \\
&&
\left[
(\vec T_q \vec T)(2\bar z z_2+(1+\varepsilon)z_3^2)+
(\vec T_{\bar q}\vec T)(z_3(2z-1)-2z\bar z+(1+\varepsilon)z_3^2)
\right]\;,
\eea
where in $(\vec T_q \vec T)$ and  $(\vec T_{\bar q}\vec T)$ the sum over the quark 
and the gluon color indices $i,j$ and $a$ is implied. The integrals over $dz_3$ 
and $d^{2+2\varepsilon}\vec q_{2}$ in (\ref{zzzz}) stand for the integration over the 
antiquark-gluon intermediate state.  For $z_3\to 0$ the integral (\ref{zzzz}) 
would diverge if the invariant mass $\kappa$ were allowed to become arbitrarily large, 
this corresponding to the radiation of the gluon in the central region, away 
from the fragmentation region of the virtual photon. Such a gluon radiation has to 
be assigned to the Reggeon Green's function and subtracted from the impact factor. 
The corresponding separation procedure, described in~\cite{FF98,FM99}, leads 
to the appearance of the $\theta(s_\Lambda - \kappa)$ in the definition of the 
impact factor (see Eq.~(\ref{13})) and to the need to subtract the so-called 
counterterm, given by the second term in the r.h.s. of Eq.~(\ref{13}). Since
\beq{slambda}
\kappa =\frac{\vec k^2}{z_3}+\frac{\vec q_2^{\,\,2}}{z_2}-\vec q^{\,\,2}
\eeq
and the integral is restricted by the condition $\theta (s_\Lambda-\kappa)$,
with $s_\Lambda\to \infty$, the lower boundary of the $z_3$ integral in
(\ref{zzzz}) is $z_3^{min}=\frac{\vec k^2}{s_\Lambda}$.

We define the divergent  part of (\ref{zzzz})
for $z_3\to 0$ as
\beq{bvbv}
-\frac{2e_qg^4f_VQ}{N_c(2\pi)^{3+2\varepsilon}}
\int\limits^{1}_0 \, \phi_\parallel (z)\, dz \int\limits^{\bar z}_0
\frac{dz_3}{z_3}d^{2+2\varepsilon}\vec q_{2}
\left[\bar z
(\vec T_q\bar z- \vec T_{\bar q}z ) \vec T
\right]_{z_3\to 0} \, , 
\eeq
where
\beq{tvtv}
\left[\bar z
(\vec T_q\bar z- \vec T_{\bar q} z) \vec T
\right]_{z_3\to 0}=\frac{N_c\delta^{cc^\prime}}{2 Q^2}
\frac{\vec q^{\,\, 2}}{\vec k^2[\vec q_2^{\,\, 2}+Q^2 z\bar z]} \, .
\eeq
Integrating over $z_3$ in (\ref{bvbv}) we obtain 
\beq{tata}
-\frac{e_qg^4f_V\delta^{cc^\prime}}{Q(2\pi)^{3+2\varepsilon}}
\int\limits^{1}_0 \, \phi_\parallel (z)\, dz
\int d^{2+2\varepsilon}\vec q_2
\frac{\vec q^{\,\, 2}}{ (\vec q_2-\vec q)^2
[\vec q_2^{\,\, 2}+Q^2 z\bar z]}\ln{\left(\frac{\bar z s_\Lambda}{(\vec
q_2-\vec q)^2}\right)} \; .
\eeq

The consideration of the contribution of the diagrams shown in Fig.~9(b)
goes along the same lines. Similarly to~(\ref{tata}), 
we define the contribution which is singular for $s_\Lambda\to \infty$, given 
by the replacement $z\to\bar z$ in~(\ref{tata}). Then in the sum of these two
contributions with the BFKL subtraction term we find that $s_\Lambda$ cancels 
out and arrive at some finite contribution to the impact factor which is denoted 
below as $T_4$.
Thus, the total contribution of the $(q\bar q g)$ intermediate state to the
NLA impact factor may be represented as the sum of two terms,
\beq{gluon}
T^{(q\bar q g)}=T_4+T_5 \, ,
\eeq
where the results for $T_4$ and $T_5$ are presented, as usual, in the form
\beq{T4,5}
T_{4,5}=\frac{g^2\Gamma [1-\varepsilon]
\left(Q^{2}/\mu^2\right)^\varepsilon}{(4\pi)^{2+\varepsilon}}
T_H^{(0)}\left[\, \tau_{4,5} (z) + \tau_{4,5}(\bar z) \, \right] \, .
\eeq
The procedure described above gives for $\tau_4$ the following expression:
\bea{tau44}
&&
\tau_4(z)=\frac{(\alpha+z\bar
z)N_c\left(Q^2\right)^{1-\varepsilon}}{\pi^{1+\varepsilon}\Gamma
[1-\varepsilon]}
\int 
\frac{d^{2+2\varepsilon}\vec q_2}{ (\vec q_2-\vec q)^2
[\vec q_2^{\,\, 2}+Q^2 z\bar z]}\ln{\left(\frac{z \bar z s_0}{(\vec
q_2-\vec q)^2}\right)} 
\nonumber \\
&&
+\frac{N_c\,
\Gamma^2[1+\varepsilon](\alpha)^{\varepsilon}}{\varepsilon 
\, \Gamma[1+2\varepsilon]}\ln{\left(\frac{\alpha Q^2}
{s_0}\right)}\; , 
\eea
and for $\tau_5(z)$ one arrives naturally at two contributions
\beq{t5at5b}
\tau_5(z)=\tau_5^a(z)+\tau_5^b(z)\;,
\eeq
where
\bea{t5a}
&&
\delta^{cc^\prime}\,\tau_5^a(z)=\frac{2\left(Q^2\right)^{1-\varepsilon}}
{\pi^{1+\varepsilon}\Gamma[1-\varepsilon]}
\frac{(\alpha+z\bar z)}{\alpha}
\int\limits^{\bar z}_0
\frac{dz_3}{z_3}d^{2+2\varepsilon}\vec q_{2}
\nonumber \\
&&
\left[2\bar z^2
\left(\vec T_q \vec T- \vec T_q \vec T|_{z_3\to 0} \right) 
-2z\bar z\left(\vec T_{\bar q} \vec T- \vec T_{\bar q} \vec T|_{z_3\to 0} 
\right)\right]
\eea
and
\bea{t5b}
&&
\delta^{cc^\prime}\,\tau_5^b(z)=\frac{2\left(Q^2\right)^{1-\varepsilon}}
{\pi^{1+\varepsilon}\Gamma[1-\varepsilon]}
\frac{(\alpha+z\bar z)}{\alpha}
\int\limits^{\bar z}_0
dz_3 d^{2+2\varepsilon}\vec q_{2}
\nonumber \\
&&
\left[\left(2z-1+z_3(1+\varepsilon)\right)
\left(\vec T_{\bar q} \vec T \right)
-\left(2\bar z-z_3(1+\varepsilon)\right)\left(\vec T_q \vec T
\right)\right] \; .
\eea

After a long calculation we find
\bea{t5aI}
&&
\tau_5^a(z)=\frac{2C_F}{\varepsilon}\int\limits_0^{\bar z}
dz_3\left(\frac{2z-1+z_3}{\alpha+z_2\bar z_2}\right)
-\frac{1}{N_c}\left(\frac{1}{\varepsilon^2}
+\frac{\ln \alpha}{\varepsilon}\right)
\nonumber \\
&&
+2C_F\left[-\frac{\pi^2}{6}+\int\limits_0^{\bar z}
dz_3\left(\frac{2z-1+z_3}{\alpha+z_2\bar z_2}\right)
\ln\left(\frac{(\alpha+z_2\bar z_2)^2}{z_2\bar z_2}\right)
\right]
\nonumber \\
&&
+N_c\int\limits_0^{\bar z}
\frac{dz_3}{z_3}\ln\left(\frac{\alpha z^2(\alpha z_2+\bar z^2\bar z_2)
(\alpha+z_2\bar z_2)^2}{\bar z\bar z_2(\alpha \bar z_2+z z_3)
(\alpha+z\bar z)^3}\right)
\nonumber \\
&&
-\frac{1}{N_c}\left[
\frac{\ln^2 \alpha}{2}+\int\limits_0^{\bar z}
dz_3\left\{
\ln(\alpha+z_2\bar z_2)\left(\frac{\bar z}{\alpha+\bar z\bar z_2}-
\frac{ z}{\alpha+z z_2}\right)
\right.\right.
\nonumber \\
&&
\left.\left.
+\frac{\bar z}{\alpha+\bar z\bar z_2}
\ln\left(\frac{\alpha z_2+\bar z^2\bar z_2}{\alpha z_3^2\bar z_2}
\right)
+\frac{z}{\alpha+ z z_2}
\ln\left(\frac{z_3(\alpha \bar z_3+z z_2)}{\alpha +z\bar z}
\right)
\right\}\right]
\eea
and
\bea{t5bI}
&&
\tau_5^b(z)=-\frac{C_F}{\varepsilon}\int\limits_0^{\bar z}
dz_3\left\{\left(
\frac{2z-1+z_3}{\alpha+z_2\bar z_2}\right)\frac{\alpha+z\bar z}{z\bar z}
-\frac{z_3-2\bar z}{\bar z^2}
\right\}
\nonumber \\
&&
+N_c\frac{\alpha+z\bar z}{\alpha}\int\limits_0^{\bar z}
dz_3 \ln\left(\frac{\alpha \bar z+z_2 z_3}{z_2 z_3}\right)
\left(
\frac{2z-1+z_3}{z\bar z}+\frac{z_3-2\bar z}{\bar z^2}
\right)
\nonumber \\
&&
+\int\limits_0^{\bar z}
dz_3\left\{C_F\left[-\frac{\alpha+z\bar z}{z\bar z}
\left(\frac{z_3}{\alpha+z_2\bar z_2}
+\frac{2z-1+z_3}{\alpha+z_2\bar z_2}\ln\left(\frac{(\alpha+z_2\bar
z_2)^2}{z_2\bar z_2}\right)\right) +\frac{z_3}{\bar z^2}
\right.\right.
\nonumber \\
&&
+\frac{2z-1+z_3}{\alpha}\ln\left(\frac{\alpha z_2+\bar z^2 \bar z_2}{\bar z
z_3}\right)
+\frac{(\alpha+z\bar z)(2z-1+z_3)}{\alpha z\bar z}
\ln\left(\frac{\bar z(\alpha \bar z_2+z z_3)(\alpha z_2+\bar zz_3)}
{z_3(\alpha+z\bar z)(\alpha z_2+\bar z^2\bar z_2)}\right)
\nonumber \\
&&
\left.
+\frac{(\alpha+z\bar z)(z_3-2\bar z)}{\bar z^2 \alpha}
\ln\left(\frac{\alpha z_2^2(\alpha z_2+\bar z z_3)}
{\bar z^3 z_3}\right)
+\frac{z(z_3-2\bar z)}{\bar z \alpha}
\ln\left(\frac{\bar z^3(\alpha \bar z_2+ z
z_3)}
{\alpha z_3 z_2^2(\alpha+z\bar z)}\right)
\right]
\nonumber \\
&&
-\frac{1}{2N_c}
\left[
\frac{(\alpha+z\bar z)(2z-1+z_3)}{z \bar z \alpha}
\left(2-\frac{z z_2}{\alpha+z z_2}-\frac{\bar z\bar z_2}{\alpha+\bar
z\bar z_2}\right)
\ln\left(\alpha + z_2\bar  z_2\right)
\right.
\nonumber \\
&&
+\frac{(\alpha+z\bar z)(z_3-2\bar z)}{\bar z^2 \alpha}
\ln\left(\frac{ z_3^3(\alpha z_3+\bar z z_2)}
{z_2^3(\alpha z_2+\bar z z_3)}\right)
+\frac{z(z_3-2\bar z)}{\bar z \alpha}
\ln\left(\frac{z_2(\alpha \bar z_3+ z
z_2)}
{ z_3 (\alpha \bar z_2+z z_3)}\right)
\nonumber \\
&&
+\frac{(\alpha+z\bar z)(2z-1+z_3)}{\alpha z\bar z}
\left(\ln\left(\frac{(\alpha+z\bar z)(\alpha z_3+\bar z z_2)(\alpha z_2+\bar
z^2\bar z_2)}
{\alpha \bar z_2 z_2^2 z_3
(\alpha \bar z_2+z z_3)(\alpha z_2+\bar z z_3)}\right)
\right.
\nonumber \\
&&
\left.\left.\left.
+\frac{z z_2}{\alpha+z z_2}
\ln\left(\frac{z_3(\alpha
\bar z_3+z z_2)}
{\alpha +z \bar z}\right)
+\frac{\bar z\bar z_2}{\alpha+\bar z \bar z_2}
\ln\left(\frac{\alpha
\bar z_2 z_3^2}
{\alpha z_2 +\bar z^2\bar z_2}\right)
\right)\right]\right\}\;.
\eea

The integration over $\vec q_2$ in $\tau_4(z)$ is quite simple and leads  
to
\[
\tau_4(z)=\left(C_F+\frac{1}{2N_c}\right)
\left[
\frac{2}{\varepsilon^2}+\frac{2\ln\alpha+4\ln z}{\varepsilon}
+4\ln\left(\frac{s_0/Q^2}{\alpha+z\bar z}\right)
\ln\left(\frac{\alpha+z\bar z}{z}\right)
\right.
\]
\beq{tau4}
+2\ln{\alpha}\ln{\left(\frac{\alpha Q^2}{s_0}\right)}-\left. 
\frac{\pi^2}{3}+\ln^2\alpha+4\ln z\ln\left(\frac{(\alpha+z\bar z)^3}{\alpha z
\bar z}\right)+2 {\rm Li}_2\left(-\frac{z\bar z}{\alpha}\right)
\right]\;.
\eeq

For $\tau_5(z)$ a more cumbersome calculation gives 
\[
\tau_5(z)=\frac{C_F}{\varepsilon}\left[\frac{\left(\alpha -z\bar z\right)}
{2\,z\bar z} \, \left( \frac{\left(1 - 2\,z \right)}{c}\,\ln(1 + 2\,c - 2\,z) 
- \ln \left(\frac{\alpha + z\bar z }{\alpha}\right) \right) -\frac{3}{2}\right]
\]
\[
+\frac{1}{N_c}\left[-\frac{1}{\varepsilon^2}-\frac{\ln\alpha}{\varepsilon}
\right]
+N_c\left[-\frac{\pi^2}{6}-\ln^2 \alpha +\ln(\alpha z)\ln(\alpha+z\bar z)
\right.
\]
\[
+\ln\left(\frac{\alpha z}{\alpha+z\bar
z}\right)\ln\left(\frac{\alpha+z}{\alpha+z\bar
z}\right)
+\frac{1}{2}\ln^2\left(\frac{\alpha z}{\alpha+z\bar z}\right)
-\frac{1}{2}\ln^2\left(\frac{z}{\bar z}\right)
\]
\[
- 
\frac{\left( \alpha + z\bar z \right)} {2\alpha \,z\,\bar z}
\left(z\,\ln\left(\frac{\alpha}{z}\right)+{\sqrt{z\,\left(4\alpha+ z\right)}}\,
\ln \left(\frac{z + {\sqrt{z\, \left( 4\alpha + z\right) }}}{-z +{\sqrt{z\,
 \left( 4\alpha + z\right) }}}\right) \right) 
+2\,{\rm Li}_2\left(\frac{2\,z}{1 - 2\,c}\right)
\]
\[
\left.
+2\,{\rm Li}_2\left(\frac{2\,z}{1 + 2\,c}
\right)+ {\rm Li}_2\left(-\frac{z\bar z}{\alpha}\right) 
+ {\rm Li}_2\left(z-\frac{\alpha}{z}\right)-{\rm Li}_2\left(-\frac{
z}{\alpha}\right) + {\rm Li}_2\left(-\frac{z^2}{\alpha + z\bar
z}\right)\right]
\]
\[
+C_F\left[3-\frac{\pi^2}{3}-2\ln \alpha \ln z
+2\ln\left(\frac{\alpha}{z}\right)\ln\left(\frac{\alpha+z\bar
z}{z}\right)- \frac{1}{2}{\ln^2 \alpha}\right.
\]
\[
- \frac{\left( 2 - z \right) \,z}{2\,\bar z \,
\left(\alpha - z \right)} + \frac{z^2}{2\,\bar z \,\left(\alpha - z^2\right) } 
-  \frac{\left( 1 - 2z \right) \left( \alpha + z\bar z \right)}{2\,
c\,z \, \bar z }\ln (1 + 2c - 2z)
\]
\[
\left.
- \left( \frac{z}{2\left(\alpha - z \right)} 
- \frac{\left( 2 - z \right) z^2}{2\bar z {\left( \alpha - z \right) }^2}
+ \frac{1 + z} {2\,\bar z }\right) \ln \left(\frac{\alpha}{z}\right) \right.
\]
\[
+ \left. {\ln^2 \left(\frac{2\,c-1} {2\,z}\right)} 
+ {\ln^2\left(\frac{1 + 2\,c}{2\,z}\right)} + 4\ln \left(\frac{1 + 2\,c 
- 2\,z}{2\,c-1}\right) \ln \left(\frac{1 + 2\,c - 2\,z} {2\,z}\right) \right.
\]
\[
-2\ln\left(\frac{1 + 2\,c}{2\,z}\right)\,\ln \left(\frac{1 + 2\,c- 2\,z}
{2\,z}\right) - \frac{\left( \alpha + z\bar z \right)}{2\,\bar z \,
\left( \alpha + z \right)}\ln z + \ln z\ln \left(\frac{1+2c-2z}{2c-1+2z}\right)
\]
\[
- \left.\left( \frac{\alpha} {2\,z\,\bar z}+\frac{z^2}{2\,\bar z\,
\left(\alpha + z \right) }\right) \, \ln \left(\frac{\alpha + z\bar z}
{\alpha}\right) + \frac{\left( \alpha + z \bar z \right)}{2\,z \,\bar z }
\left(\frac{{\ln^2 \alpha}}{2} + 2\ln z\,\ln \left(\frac{\alpha + z\bar z}
{\alpha}\right)\right.\right.
\]
\[
+ \left.\left. \ln^2 \left(\frac{2\,c-1}{2}\right) + \ln^2 \left(\frac{1+2\,c}
{2}\right) - {\ln^2 (\alpha + z\bar z)}  \right) 
- \ln \left(\frac{1 + 2\,c}{2\,z}\right)\, \ln \left(\frac{\alpha + z\bar z}
{z^2}\right)\right.
\]
\[
- \left.\left(\frac{z^3}{2\,\bar z \,{\left(\alpha - z^2\right)}^2}
+ \frac{z}{\alpha - z^2}+\frac{5 - 7\,z}{2\,\bar z }\right) \,
\ln \left(\frac{\alpha + z\bar z }{z}\right) 
 \right.
\]
\[
- \left.\frac{\left(1-2\,z \right) \,\left(\alpha - z\bar z\right)\,}{2\,c\,z\,
\bar z} \left(\ln z\,\ln \left(\frac{2\alpha + \left( 1 - 2\,c \right) \,z}
{2\alpha + \left( 1 + 2\,c \right) \,z}\right)  - \ln (4\alpha+1)\,
\ln \left(\frac{1 + 2\,c - 2\,z}{2\,c -1 + 2\,z}\right) \right.\right.
\]
\[
+ \left.\left.2\,{\rm Li}_2\left(\frac{1 - 2\,c - 2\,z}{1 + 2\,c - 2\,z}\right)
\right)+ \frac{\left(1+2\,c - 2\,z \right)\,\left( \alpha - z \bar z 
\right) }{2\,c\,z\bar z} \, {\rm Li}_2\left(\frac{2\,z}{1 - 2\,c}
\right) \right.
\]
\[
+ \left.  \frac{\left(2c -1 + 2z \right) \,\left( \alpha -z\bar z
\right) }{2\,c\,z\,\bar z}  {\rm Li}_2\left(\frac{2\,z}{1 + 2c}  
\right) \right]
\]
\[
+\frac{1}{N_c}\left[ 
-\frac{1}{2\,z} - \frac{2\,z - z^2} {2\,\bar z \,
\left( \alpha - z \right) } + \frac{\left( \alpha + z\bar z  \right)}
{z\,\bar z}\ln\alpha + \frac{1}{2}\ln \left(\frac{{\left(2\,c-1 \right)}^2}{4\,
\alpha}\right)\,\ln \left(\frac{ 2\,c-1}{2\,c+1}\right) \right.
\]
\[
+ \left.\frac{c\left(1 - 2z \right)\left(\alpha + z\bar z \right)}{z^2\bar z^2}
\ln (1 + 2c - 2z) + \frac{\alpha\left( 1 - 2z \right)\left( \alpha + z\bar z
\right) }{4\,z^3\bar z^3}\ln 2\ln (1 + 2c - 2z) \right.
\]
\[
+ \left.\left(\frac{7}{4}-\frac{\alpha\,\left(2 - z\right)}{4 z^2\,\bar z^2} 
- \frac{\alpha^2\, \left(2 - 4\,z + 3\,z^2 \right) }{4\, z^3\,\bar z^3} \right)
{\ln^2 (1 + 2\,c - 2\,z)} \right.
\]
\[
+ \left. \left( \frac{z} {2\bar z \left( \alpha - z \right) } 
+ \frac{2z^2 - z^3} {2\,\bar z \, {\left( \alpha - z \right) }^2} \right) 
\ln \left(\frac{\alpha}{z}\right)-\ln (1 + 2c - 2z)\,\ln ( 2c-1 + 2z)\right.
\]
\[
- \left.\left(\frac{\alpha^2}{z^3\,\bar z^3} + \frac{\alpha - 2\alpha^2}
{z^2\,\bar z^2} \right) \ln \alpha\ln z 
- \frac{\left( \alpha + z\bar z \right)}{4\,z^2\,\bar z^2}\ln (\alpha 
+ z\bar z)\right.
\]
\[
+ \left.\left(\frac{3 - 4\,z}{2\,\bar z }+\frac{\alpha\,\left(1- 2\,z \right) }
{z\,\bar z^2} + \frac{z}{2\,\bar z \,   \left( \alpha - z^2 \right) }\right) 
\ln \left(\frac{z}{\alpha + z\bar z}\right)   
\right.
\]
\[
- \left. \frac{\left( \alpha^2 + \alpha z - \bar z z^3\right)}
{2\,\bar z \,z^3} \left(-\frac{{\pi }^2}{6} - 7\,{\ln^2 2} - 2\ln 2\ln \alpha 
- \frac{1}{2}{\ln^2 \alpha} + {\rm Li}_2\left(-\frac{z^2}{\alpha + z\bar z}  
\right)\right.\right.
\]
\[
+ \left. {\ln^2 (2\,c-1)} + {\ln^2 (2\,c+1)} - \frac{1}{2}{\ln^2 z} -  \ln 2
\ln (2\,c-1 + 2\,z) - 2 {\rm Li}_2\left(- \frac{\bar z z}{\alpha}\right)\right.
\]
\[
+ 4\ln (1 + 2c - 2z) \ln (2c -1+ 2z) + \frac{1}{2}{\ln^2 ( 2c-1 + 2z)} +    
\ln \left(\frac{\alpha + z}{\alpha}\right) \ln \left(\frac{\alpha\,z}
{\alpha + z\bar z}\right) 
\]
\[
- \frac{11}{2}\ln 2\,\ln (\alpha + z\bar z)-2\ln \alpha \ln (\alpha + z\bar z) 
- 2\ln \bar z\,\ln (\alpha + z\bar z) - \ln z \ln (\alpha + z\bar z) 
\]
\beq{tau5}
- \left.\left. {\rm Li}_2\left(- \frac{z}{\alpha} \right) 
+ 2 {\rm Li}_2\left(\frac{2\,z}{1 - 2\,c}\right) 
+ 2 {\rm Li}_2\left(\frac{2\,z}{1 + 2\,c}\right)  
- {\rm Li}_2\left(z - \frac{\alpha}{z}\right) \right)
\right]\;.
\eeq
Equations (\ref{gluon}), (\ref{T4,5}), (\ref{tau4}) and (\ref{tau5}) give 
the result for the $(q\bar qg)$ contribution to the impact factor.

\subsection{The result for NLA impact factor}

Now we are ready to add together the $(q\bar q)$ and the $(q\bar qg)$ contributions
and to obtain (as anticipated above), after the treatment of the ultraviolet 
and of the collinear singularities, the finite result for the NLA impact factor.   

In the sum of all contributions -- the $z\leftrightarrow \bar z$ symmetry
was used where necessary -- we find
\beq{sumtau}
\sum_{i=1,\dots,5}\tau_i(z)= \frac{C_F}{\varepsilon}\left[\frac{3}{2}  
+ \frac{\left(1-2\,z\right)\left(\alpha-\bar z z\right)}{4\,c\, \bar z \,z}\,
\ln \left(\frac{1 + 2\,c - 2\,z}{2\,c -1+ 2\,z}\right)
- \frac{\left( \alpha + \bar z z \right)}{4\,c\,\bar z \, z} 
\ln \left(\frac{2\,c+1}{2\,c-1}\right) \right.
\eeq
\[
- \left. \frac{\left(\alpha - \bar z z \right)}{2\, \bar z \, z}
\ln\left(\frac{\alpha+\bar z z}{\alpha}\right)\right] 
-\frac{\beta_0}{2\,\varepsilon} + \dots \; ,
\]
where the ellipsis stand for the terms which are finite for $\varepsilon\to 0$.
First of all, we note that the double poles in $\varepsilon$ which appear 
due to soft singularities and are present in the separate contributions cancel 
out in the sum~(\ref{sumtau}).
What is left are the single poles in $\varepsilon$ which represent the 
ultraviolet and the collinear singularities which are removed by 
the renormalization of the strong coupling constant and of the distribution 
amplitude, as explained in~Section 2.

After substituting the bare strong coupling constant and the meson distribution 
amplitude by the renormalized quantities, given respectively in 
Eqs.~(\ref{alsrem}) and (\ref{renampl}), we find the following expressions for 
the ultraviolet
\beq{UVC}
\Delta^{\alpha_s}\, T_H(z)=\frac{\alpha_s(\mu_R)}{4\pi}\beta_0
\left[
\frac{1}{\hat \varepsilon}+\ln\left(\frac{\mu_R^2}{\mu^2}
\right)\right]T^{(0)}_H(z)\, ,
\eeq
and the collinear 
\beq{COC}
\Delta^{coll}\, T_H(z)=
-\frac{\alpha_s(\mu_F)}{2\pi}
\left[\frac{1}{\hat
\varepsilon}+\ln\left(\frac{\mu_F^2}{\mu^2}\right)\right]
\int\limits^1_0 du \, T^{(0)}_H(u)\,  V^{(1)}(u,z) \; ,
\eeq
counterterms to $T_H$, where the LLA hard-scattering amplitude $T^{(0)}$  
and the one-loop evolution kernel $V^{(1)}$ are given by Eqs.~(\ref{LO})  
and~(\ref{V(1)}), respectively. Calculating the integral in (\ref{COC}) and 
then combining the corresponding contributions coming from the counterterms
with the singular part of (\ref{sumtau}), we see that the ultraviolet and the
collinear singularities in the hard scattering amplitude cancel and 
the result for the finite part may be presented as in~(\ref{sumtau})
with the replacements
\beq{mus}
\frac{\beta_0}{\varepsilon}\to\beta_0\ln\left(\frac{Q^2}{\mu_R^2}\right)\
, \;\;\;\;\;
\frac{C_F}{\varepsilon}\to C_F\ln\left(\frac{Q^2}{\mu_F^2}\right) \, .
\eeq

Finally we find that the NLA impact factor is given by Eq.~(\ref{fact}) with 
\beq{pppp}
T_H(z,\alpha,s_0,\mu_F,\mu_R)|_{\alpha\to 0}=
\alpha_S(\mu_R)\frac{\alpha}{\alpha +z\bar z}\left\{
1+\frac{\alpha_s(\mu_R)}{4\pi}\left[\tau(z)+\tau(\bar z)\right]
\right\} \; ,
\eeq
where
\[
\tau(z)= C_F\ln\left(\frac{Q^2}{\mu_F^2}\right)
\left[\frac{3}{2}  
+ \frac{\left(1-2\,z\right)\left(\alpha-\bar z z\right)}{4\,c\, \bar z \,z}\,
\ln \left(\frac{1 + 2\,c - 2\,z}{2\,c -1+ 2\,z}\right)
- \frac{\left( \alpha + \bar z z \right)}{4\,c\,\bar z \, z} 
\ln \left(\frac{2\,c+1}{2\,c-1}\right) \right.
\]
\[
- \left. \frac{\left(\alpha - \bar z z \right)}{2\, \bar z \, z}
\ln\left(\frac{\alpha+\bar z z}{\alpha}\right)\right] 
-\frac{\beta_0}{2}\ln\left(\frac{Q^2}{\mu_R^2}\right) + n_f\left[-\frac{5}{9}+\frac{\ln \alpha}{3} 
\right] + C_F\left[-\frac{{\ln^2\alpha}}{2}-3\ln\left(\frac{\alpha+\bar z z} 
{z}\right)\right.
\]
\[
+\left.4\,{\ln^2 \left(c-z+\frac{1}{2}\right)}-2\ln\left(c-\frac{1}{2} \right)
\ln \left(c+\frac{1}{2} \right) + \frac{\left( \alpha + \bar z z \right)}
{2\,\alpha\, \bar z \,z} \left(\ln \alpha + 2\,c \ln \left(\frac{2\,c+1}    
{2\,c-1}\right) \right) \right.
\]
\[
- \left. \frac{\left(1 - 2\,z \right) \left( \alpha + \bar z z \right)}
{2\,c\,\bar z \,z}\ln (1 + 2\,c - 2\,z)-\frac{\left(\alpha + \bar z z \right)}
{2\,\bar z \,z} \ln \left(\frac{\alpha + \bar z z }{\alpha}\right)
+ \frac{\left( \alpha + \bar z z \right)} {\bar z \, \left( \alpha + z \right)}
\ln \left(\frac{\alpha + \bar z z }{\alpha\,z}\right)  \right.
\]
\[
+ \left. \frac{\left(\alpha + \bar z z \right)}{2\,c\, \bar z \,z}
\left(\frac{{\pi }^2}{6}+\ln 4\ln\alpha+\frac{1}{2}\ln\left(\frac{1+2\,c}
{2\,c-1}\right)+2\ln c \ln \left(\frac{2\,c-1} {4}\right) - {\ln^2 (2\,c+1)} 
\right.\right.
\]
\[
- \left.\left.{\ln^2 c} +\frac{1}{4}\ln^2 \left(\frac{2\,c+1}{2\,c-1}\right)
- 2{\rm Li}_2\left(\frac{2\,c-1}{4\,c}\right) 
- {\rm Li}_2\left(\frac{2}{1 + 2\,c}\right) \right)
+ \frac{\left(\alpha + \bar z z \right)} {2\,\bar z \,z}
\left( \frac{{\ln^2 \alpha}}{2} \right.\right.
\]
\[
- \left.\left. 2\ln \left(c-\frac{1}{2}\right)\ln \left(c+\frac{1}{2} \right)
+ {\ln^2 \left(\frac{z}{\alpha}\right)} - {\ln^2 \left(\frac{\alpha 
+ \bar z z}{z}\right)}+ 2{\rm Li}_2\left(\frac{2\,z}{1 - 2\,c}\right)
\right.\right.
\]
\[
+ \left. 2{\rm Li}_2\left(\frac{2\,z}{1 + 2\,c}\right) \right) 
- \frac{\left(1 - 2\,z \right)\left(\alpha -\bar z z \right) } 
{2\,c\, \bar z \,z}\left(\ln \left(\frac{z} {4\alpha+1}\right) 
\ln \left(\frac{2\alpha + \left( 1 - 2\,c \right) z}{2\alpha 
+ \left( 1 + 2\,c \right) z}\right)\right.  
\]
\[
- \left. \ln (4\alpha+1)\, \ln \left(\frac{2\,c+1} {2\,c-1}\right) 
+ 2{\rm Li}_2\left(\frac{1 - 2\,c - 2\,z}{1 + 2\,c - 2\,z} \right) 
- {\rm Li}_2\left(\frac{2\,z}{1 - 2\,c}\right) \right. 
\]
\[
+ \left.\left. {\rm Li}_2\left(\frac{2\,z} {1 + 2\,c}\right) \right)\right]  
+ N_c \left[ \ln \left(s_0/Q^2\right)\,\ln \left(\frac{(\alpha + \bar z z)^2
}{z^2\alpha}
\right) +\frac{20}{9}  - \frac{\ln \alpha}{3} 
+ \frac{1}{2}\ln^2\left(\frac{\bar z}{z}\right) \right.
\]
\[
 \left. +\frac{1}{2}\ln^2 \left(\frac{\alpha + \bar z
z}{\alpha}\right)  
-2\ln \left(\frac{\alpha+z}{z}\right)\ln \left(\frac{\alpha + \bar z
z}{\alpha\ z}\right)-\ln^2 \left(\frac{\alpha + \bar z
z}{z}\right)+3\, \ln z\, 
\ln \left(\frac{ \alpha + \bar z z }{\alpha \,z}
\right) \right.
\]
\[
+2\, \ln^2 z
- \left.\frac{\left(\alpha+ \bar z z \right)} {2\,\alpha\,\bar z \,z}
\left(z\ln\left(\frac{\displaystyle \alpha}{\displaystyle z}\right) 
+ {\sqrt{z \left( 4\,\alpha + z \right) }} 
\ln \left(\frac{z + {\sqrt{z\,\left( 4\,\alpha + z \right) }}}{-z + 
{\sqrt{z\, \left( 4\,\alpha + z \right) }}}\right) \right) 
\right.
\]
\[
+ \left. {\rm Li}_2\left(\frac{2\,z}{1 - 2\,c}\right) 
+ {\rm Li}_2\left(\frac{2\,z}{1 + 2\,c}\right)
+2{\rm Li}_2\left(- \frac{z^2}{\alpha+\bar z z} \right)
- 2{\rm Li}_2\left(- \frac{z}{\alpha} \right) 
+ 3\, {\rm Li}_2\left(-\frac{\bar z z}{\alpha}\right)  \right] 
\]
\[
+ \frac{1}{N_c}\left[\frac{5}{2} + \left(\frac{\alpha + \bar z z}{\bar z \,z} 
-\frac{3}{2}\right) \ln \alpha + \frac{1}{2}
\ln \left(\frac{4\,\alpha}{{\left(2\,c-1\right) }^2}\right)
\ln\left(\frac{2\,c+1} {2\,c-1}\right)
+ \frac{1}{2}\ln^2 \left(\frac{1 + 2\,c - 2\,z}{2\,c -1+ 2\,z}\right)\right.
\]
\[
+ \left. \frac{c \left(1 - 2\,z \right) \left(\alpha + \bar z z \right)}
{{\bar z }^2\,z^2} \ln (1 + 2\,c - 2\,z)
+\frac{\left(1 - 2\,z \right)\left( \alpha +  \bar z z \right)}{\bar z^2 z} 
\ln \left(\frac{z} {\alpha+\bar z z}\right)
+ {\rm Li}_2\left(\frac{2\,z}{1 + 2\,c}\right)\right.
\]
\[
+ \left. \ln z\ln \left(\frac{\alpha + \bar z z}{\alpha}\right) 
+ \frac{\left(\alpha + \bar z z \right)}{4\,{\bar z}^2z^2} 
\,\left( 2\,c\, \ln \left(\frac{2\,c+1}{2\,c-1}\right)
-\ln\left(\frac{\alpha+\bar z z}{\alpha}\right)
\right) + {\rm Li}_2\left(\frac{2\,z}{1 - 2\,c}\right)  \right.
\]
\[
- \left. \frac{\left( \alpha^2 + \alpha\,z - \bar z \,z^3 \right)}
{2\,\bar z\,z^3}\left(2\ln \left(c-z +\frac{1}{2}\right)\ln \left(c+z- \frac{1}
{2} \right)-2  \ln \left(c- \frac{1}{2}\right)\ln \left(c+\frac{1}{2}\right)
\right. \right.
\]
\[
+ \left. \ln \left(\frac{2\,c+1}{2\,c-1}\right)\ln \left(\frac{2\,\alpha    
+ \left( 1 + 2\,c\right) \,z}{2\,\alpha + \left( 1 - 2\,c\right) \,z}\right) 
- \ln \left(\frac{\alpha\,{\bar z}^2\,z^2}{{\left(\alpha +\bar z z \right) }^2}
\right) \ln \left(\frac{\alpha + \bar z z}{\alpha}\right) \right.
\]
\[
+2\ln \left(\frac{\alpha + z}{\alpha}\right)\ln \left(\frac{\alpha\,z}
{\alpha + \bar z z }\right) - 2{\rm Li}_2\left(-\frac{z}{\alpha} \right)    
+  4{\rm Li}_2\left(\frac{2\,z}{1 - 2\,c}\right) + 4{\rm Li}_2\left(\frac{2\,z}
{1 + 2\,c}\right)
\]
\beq{sumtau1}
- \left.\left. 2{\rm Li}_2\left(-\frac{\bar z z}{\alpha}\right)
+ 2{\rm Li}_2\left(-\frac{z^2}{\alpha + \bar z z}  \right) \right)\right]\;.
\eeq
Equations (\ref{fact}), (\ref{pppp}) and (\ref{sumtau1})
represent the main result of the present paper.

In LLA terms of the type $\sim \alpha_s^n \ln^n(s/s_0)$ are summed.
Therefore the leading energy scale uncertainty of a LLA amplitude
is related with the contributions $\sim \alpha_s^n
\ln^{n-1}(s)\ln(s_0)$. These terms represent some particular part of
NLA correction to the amplitude. Note that in the sum of all NLA terms
the contributions $\sim \alpha_s^n
\ln^{n-1}(s)\ln(s_0)$ should cancel since
by construction a NLA amplitude does not depend 
on $s_0$ with the NLA accuracy.

Using our result for the meson impact factor let us demonstrate
the cancellation of the  $\sim \alpha_s^n
\ln^{n-1}(s)\ln(s_0)$ contributions to the forward amplitude of the process
of two $\rho$ mesons production,
$\gamma_1^*(Q_1^2)\gamma_2^*(Q_2^2)\to \rho_1 \rho_2$. Examining
Eq.~(\ref{11}) one can see that the contributions of such type
can not be generated by the corrections to the kernel of BFKL equation,
since these corrections do not depend on $s_0$. Therefore, for our purposes,
we can use the known result for the Green's function in the LLA approximation.
In the azimuth symmetric, $n=0$, sector it reads
\beq{GreenF}
G_\omega(\vec q_1, \vec q_2, \vec \Delta =0)=\int\limits^{+\infty}_{-\infty}
\frac{d\nu}{2\pi^2}\frac{(\vec q_1^{\,\, 2})^{\frac{1}{2}+i\nu}(\vec
q_2^{\,\, 2})^{\frac{1}{2}-i\nu}}{\omega-\bar \alpha_S\chi (\nu)} \,
\eeq
where
\beq{chii}
\bar \alpha_S=\frac{\alpha_S N_c}{\pi}\, , \quad
\chi(\nu)=
2\Psi (1)
-\Psi\left(\frac{1}{2}+i\nu\right)-\Psi\left(\frac{1}{2}-i\nu\right)
 \, ,
\eeq
and $\Psi(x)=\Gamma'(x)/\Gamma(x)$. Inserting (\ref{GreenF}) in Eq.~(\ref{11})
we found the following representation for the
amplitude
\beq{xxxv}
{\cal I}m_s ({\cal A})_{\gamma_1^*\gamma_2^*}^{\rho_1\rho_2}=
\frac{s}{(2\pi)^2}\int\limits^{+\infty}_{-\infty}d\nu
\left(\frac{s}{s_0}\right)^{\bar \alpha_S
\chi(\nu)}\, I_1(\nu, s_0)I_2(\nu, s_0) \, ,
\eeq
here
\bea{II11}
&&
I_1(\nu, s_0)=\int d^2 q_1 \frac{(\vec q_1^{\,\,
2})^{i\nu-\frac{3}{2}}}{\sqrt{2}\pi}\Phi_{\gamma_1^*\to\rho_1}(\vec
q_1^{\,\,
2}, s_0) \, ,
\nonumber \\
&&
I_2(\nu, s_0)=\int d^2 q_2 \frac{(\vec q_2^{\,\,
2})^{-i\nu-\frac{3}{2}}}{\sqrt{2}\pi}\Phi_{\gamma_2^*\to\rho_2}(\vec
q_2^{\,\,
2}, s_0) \, .
\eea
According to our calculation
\bea{dsds}
&&
\Phi_{\gamma^*_L\to\rho_L}\left(\alpha=\frac{\vec q^{\,\, 2}}{Q^2},
s_0\right)=
\nonumber \\
&&
-\frac{4\pi e f_\rho \delta^{cc^\prime}}{\sqrt{2}N_c
Q}\int\limits^1_0 dz \, \phi_\parallel (z)\,
\alpha_s \frac{\alpha}{\alpha +z\bar z}\left[
1+\frac{\bar \alpha_S }{4}\ln(s_0)\ln\left(\frac{(\alpha+z\bar z)^4}{\alpha^2z^2\bar z^2}
\right)
+\dots
\right]\, ,
\eea
the ellipsis in (\ref{dsds})
stand for the NLA contributions to the impact factor which
do not depend on $s_0$. Performing the integration over Reggeon momenta
in Eqs.~(\ref{II11}) we found that
\bea{II22}
&&
I_{1,2}(\nu, s_0)\sim (Q_{1,2}^2)^{\pm i\nu-1}J(\pm \nu)\int\limits^\infty_0
dy\frac{y^{\pm i\nu-\frac{1}{2}}}{y+1}\left[
1+\frac{\bar \alpha_S}{2}\ln(s_0)\ln\left(\frac{(y+1)^2}{y}\right)+\dots
\right]
\nonumber \\
&&
= (Q_{1,2}^2)^{\pm i\nu-1}J(\pm \nu)\frac{\pi}{\cosh(\pi \nu)}
\left[
1+\frac{\bar \alpha_S\chi(\nu)}{2}\ln(s_0)+\dots
\right]\, ,
\eea
where
\beq{II32}
J(\nu)=\int\limits^1_0 dz \, \phi_\parallel
(z)\left(z\bar z\right)^{i\nu-\frac{1}{2}}\, .
\eeq
Thus we obtain that
\bea{II33}
&&
{\cal I}m_s ({\cal A})_{\gamma_1^*\gamma_2^*}^{\rho_1\rho_2}\sim
\frac{s f^2_\rho}{Q_1^2Q_2^2}\int\limits^{+\infty}_{-\infty}d\nu
J(\nu)J(-\nu)\left(\frac{Q_1^2}{Q_2^2}\right)^{i\nu}\frac{\pi^2}{\cosh^2(\pi
\nu)}
\nonumber \\
&&
\times \left(\frac{s}{s_0}\right)^{\bar \alpha_S\chi(\nu)}
\left[1+\frac{\bar \alpha_S\chi(\nu)}{2}\ln(s_0)+\dots \right]^2 \, .
\eea
Expanding in series the last line of (\ref{II33}), one can easily see
that the dependence of the amplitude on $s_0$ indeed cancels with the NLA
accuracy. This calculation shows that the $s_0$ dependence of our result
for the impact factor is consistent with the form of the BFKL kernel.

The hard-scattering amplitude in the limits of small
and large virtualities of the Reggeized gluon reads
\[
T_H(z,\alpha,s_0,\mu_F,\mu_R)|_{\alpha\to 0}=
\alpha_S(\mu_R)\frac{\alpha}{z\bar z}\left[1+
\frac{\alpha_S(\mu_R)}{4\pi}\left(
n_f\left[-\frac{10}{9}+\frac{2\ln \alpha}{3}\right]
\right.\right.
\]
\[
-\beta_0\ln\left(\frac{Q^2}{\mu_R^2}\right)+N_c\left[
\frac{67}{9}+\frac{\ln\alpha}{3}-\ln^2 \alpha
-3\ln(z\bar z)+2\ln\left(\frac{s_0}{Q^2}\right)\ln \left(\frac{z\bar
z}{\alpha}\right)
\right.
\]
\[
\left.
-\frac{5\pi^2}{6}
+\ln^2 z+\ln^2 \bar z +\ln z \ln \bar z \,\, \right]
+C_F
\left[2z\ln \bar z+2\bar z\ln z-\frac{\pi^2}{3}
\right.
\]
\[
+z\ln^2 \bar z+\bar z\ln^2 z-8
-2\ln \alpha+2\, z \, {\rm Li}_2(z)+2\,\bar z\, {\rm Li}_2(\bar z)
\]
\beq{smallall}
\left.\left.\left.
+\left(3+2z\ln\bar z+2\bar z\ln z\right)\ln\left(\frac{Q^2}{\mu_F^2}
\right)
\right]\right)\right] +{\cal O}\left(\alpha_S^2(\mu_R)\, \alpha^2\right)\;;
\label{altozero}
\eeq

\[
T_H(z,\alpha,s_0,\mu_F,\mu_R)|_{\alpha\to \infty}=
\alpha_S(\mu_R)\left[1+
\frac{\alpha_S(\mu_R)}{4\pi}\left(
n_f\left[-\frac{10}{9}+\frac{2\ln \alpha}{3}\right]
\right.\right.
\]
\[
-\beta_0\ln\left(\frac{Q^2}{\mu_R^2}\right)+N_c\left[
\frac{22}{9}-\frac{20\ln\alpha}{3}-2\ln^2 \alpha
-\frac{\ln\bar z}{2\, z}-\frac{\ln z}{2\, \bar z}
+3\ln (z\bar z)
\right.
\]
\[
\left.\left.\left.
+2\ln\left(\frac{s_0}{Q^2}\right)\ln \left(\frac{\alpha}{z\bar z}\right)
+4\ln \alpha\ln (z\bar z)-2\ln z\ln \bar z-3\ln^2 z-3\ln^2 \bar z \right]\right)
\right]
\]
\beq{largeal}
+\, {\cal O}\left(\frac{\alpha_S^2(\mu_R)}{\alpha}\right) \;.
\label{altoinfty}
\eeq
These limiting expressions 
may be useful for the analysis of vector meson production
amplitudes in the asymmetric kinematics, when the two virtualities
are essentially different, $Q_1\gg Q_2$. In this case
not only large $\ln(s)$, but also large $\ln(Q_1^2/Q_2^2)$ appear 
in the perturbative expansion. In the BFKL approach  powers of  
the logarithm of virtuality originate from the region of Reggeon
momenta where $Q_1^2\gg \vec q^{\,\, 2}\gg Q_2^2$, 
which corresponds to the limits
(\ref{altozero}), (\ref{altoinfty}) for the meson impact factor.

\section{Summary and conclusions}

In this paper we have considered the impact factor for the virtual photon to
light vector meson transition in the case of the dominant
longitudinal polarization.
We have shown that, up to power suppressed corrections,
both in the leading and in the next-to-leading approximation
the expression for the impact factor factorizes into the convolution of a
perturbatively calculable hard-scattering amplitude $T_H$ and a meson twist-2
distribution amplitude. We have determined $T_H$ in the next-to-leading order
by calculating cut diagrams with effectively no more than two-particle
intermediate states. We have observed cancellation of the soft infrared
divergences between the ``real'' and the ``virtual'' parts of the
radiative corrections. Collinear infrared divergences have been absorbed
into the definition of the distribution amplitude. We have obtained finally
a close analytical expression for the impact factor which can be used,
after convolution with the Green's function of two interacting Reggeized
gluons, to build entirely within perturbative QCD the complete amplitude
for a physical process, in the next-to-leading logarithmic approximation.

This result can have important theoretical implications, since it could shed light
on the correct choice of energy scales in the BFKL approach and could be used
to compare different approaches such as BFKL and DGLAP. Moreover, it can be
considered as the first step towards the study of
phenomenologically relevant processes, such as the vector meson electroproduction
at the HERA collider and the production of two mesons in the photon collision
which can be studied at high-energy $e^+e^-$ and $e\gamma$ colliders.

{\it This work was completed after M.I.~Kotsky tragically passed away. 
We had the pleasure to know him not only as a bright scientist, but also
as an exceptionally rich personality. We missed a friend.}

{\large \bf Acknowledgments} 
We thank V.S. Fadin for many stimulating discussions. 
D.I. thanks the Dipartimento di Fisica dell'Universit\`a della Calabria 
and the Istituto Nazionale di Fisica Nucleare (INFN), Gruppo collegato di Cosenza,
for the warm hospitality while part of this work was done and for the
financial support. The work of D.I. was supported also by the Alexander von 
Humboldt Foundation and by RFBR 03-02-17734.


\begin{thebibliography}{99}

\bibitem{BFKL}
V.S.~Fadin, E.A.~Kuraev, L.N.~Lipatov, Phys. Lett. {\bf B60} (1975) 50;
E.A.~Kuraev, L.N.~Lipatov and V.S.~Fadin, Zh. Eksp. Teor. Fiz. {\bf 71} (1976)
840 [Sov. Phys. JETP {\bf 44} (1976) 443]; {\bf 72} (1977) 377 [{\bf 45} (1977)
199]; Ya.Ya.~Balitskii and L.N.~Lipatov, Sov. J. Nucl. Phys. {\bf 28} (1978)
822.

\bibitem{CDR98}
A.M.~Cooper-Sarkar, R.C.E.~Devenish and A.~De Roeck,
Int.~J.~Mod.~Phys.~{\bf A13} (1998) 3385, and references therein.

\bibitem{FF98}
V.S.~Fadin, R.~Fiore, Phys. Lett. {\bf B440} (1998) 359.

\bibitem{LF89}
L.N.~Lipatov, V.S.~Fadin, Sov. J. Nucl. Phys. {\bf 50} (1989) 712.

\bibitem{FRK95}
V.S.~Fadin, R.~Fiore, M.I.~Kotsky, Phys. Lett. {\bf B359} (1995) 181.

\bibitem{FRK96}
V.S.~Fadin, R.~Fiore, M.I.~Kotsky, Phys. Lett. {\bf B387} (1996) 593.

\bibitem{FL93}
V.S.~Fadin, L.N.~Lipatov, Nucl. Phys. {\bf B406} (1993) 259.

\bibitem{FFFKLQ}
V.S.~Fadin, R.~Fiore, A.~Quartarolo, Phys. Rev. {\bf D50} (1994) 5893; 
V.S.~Fadin, R.~Fiore, M.I.~Kotsky, Phys. Lett. {\bf B389} (1996) 737; 
V.S.~Fadin, L.N.~Lipatov, Nucl. Phys. {\bf B477} (1996) 767; V.S.~Fadin,    
M.I.~Kotsky, L.N.~Lipatov, Phys. Lett. {\bf B415} (1997) 97; V.S.~Fadin,    
R.~Fiore, A.~Flachi, M.I.~Kotsky, Phys. Lett. {\bf B422} (1998) 287.

\bibitem{CCH}
S.~Catani, M.~Ciafaloni, F.~Hautman, Phys. Lett. {\bf B242} (1990) 97; Nucl.
Phys. {\bf B366} (1991) 135; G.~Camici and M.~Ciafaloni, Phys. Lett. {\bf
B386} (1996) 341; Nucl. Phys. {\bf B496} (1997) 305.

\bibitem{FL98}
V.S.~Fadin, L.N.~Lipatov, Phys. Lett. {\bf B429} (1998) 127.

\bibitem{CC98}
G.~Camici and M.~Ciafaloni, Phys. Lett. {\bf B430} (1998) 349.

\bibitem{FFKP99}
V.S.~Fadin, R.~Fiore, M.I.~Kotsky and A.~Papa, Phys. Rev. {\bf
D61} (2000) 094005; Phys. Rev. {\bf D61} (2000) 094006.

\bibitem{Cia}
M.~Ciafaloni and D.~Colferai, Nucl. Phys. {\bf B538} (1999) 187.
\bibitem{Bartels:2002yj}
J.~Bartels, D.~Colferai and G.~P.~Vacca,
Eur.\ Phys.\ J.\ C {\bf 29} (2003) 235

\bibitem{Bartels:2002uz}
J.~Bartels, D.~Colferai, S.~Gieseke and A.~Kyrieleis,
Phys.\ Rev.\ {\bf D66} (2002) 094017

\bibitem{Fadin:2002tu}
V.S.~Fadin, D.Yu.~Ivanov and M.I.~Kotsky,
Nucl.\ Phys.\ {\bf B658} (2003) 156

\bibitem{FM99} 
V.S.~Fadin and A.D.~Martin, Phys. Rev. {\bf D60} (1999) 114008.

\bibitem{earlyCZ}
V.L.~Chernyak and A.R.~Zhitnitsky, JETP Lett. {\bf {25}} (1977) 510;
Yad. Fiz. {\bf 31} (1980) 1053;
V.L.~Chernyak, V.G.~Serbo and A.R.~Zhitnitsky, JETP Lett. {\bf
26} (1977) 594; Sov. J. Nucl. Phys. {\bf 31} (1980) 552.

\bibitem{earlyBL}
G.P.~Lepage and S.J.~Brodsky, Phys. Lett. {\bf B87} (1979) 359;
Phys. Rev. Lett. {\bf 43} (1979) 545, 1625 (E);
Phys. Rev. {\bf D22} (1980) 2157;
S.J.~Brodsky, G.P.~Lepage and A.A.~Zaidi, Phys. Rev. {\bf D23} (1981) 1152.

\bibitem{earlyER}
A.V.~Efremov and A.V.~Radyushkin, Phys. Lett. {\bf B94} (1980)
245; Teor. Mat. Fiz. {\bf {42}} (1980) 147.

\bibitem{Braun}
P.~Ball, V.M.~Braun, Y.~Koike and K.~Tanaka,
Nucl. Phys. {\bf B529} (1998) 323.

\bibitem{IG}
I.F.~Ginzburg and D.Yu.~Ivanov,
Phys. Rev. {\bf D54} (1996) 5523.

\bibitem{Ivanov:1998gk}
D.Yu.~Ivanov and R.~Kirschner,
Phys.\ Rev.\ {\bf D58} (1998) 114026.

\bibitem{Ivanov:2000uq}
D.Yu.~Ivanov, R.~Kirschner, A.~Schafer and L.~Szymanowski,
Phys.\ Lett.\ {\bf B478}, 101 (2000)
[Erratum-ibid.\ {\bf B498}, 295 (2001)].
   
\bibitem{FIK}
V.S.~Fadin, D.Yu.~Ivanov and M.I.~Kotsky,
Phys. Atom. Nucl.  {\bf 65} (2002) 1513
[Yad. Fiz.  {\bf 65} (2002) 1551].

\bibitem{FFQ94}
V.S.~Fadin, R.~Fiore, A.~Quartarolo, Phys. Rev. {\bf D50} (1994) 2265.

\end{thebibliography}
\end{document}